\documentclass[12pt]{iopart}
\usepackage{iopams} 
\usepackage{mathbbol} 
\usepackage[colorlinks, linkcolor =blue]{hyperref} 
\usepackage{amsthm} 
\usepackage{mathrsfs} 
\usepackage{array} 
\usepackage{multirow} 
\usepackage{color}

\newcommand{\bm}[1]{\boldsymbol{#1}} 
\newcommand{\binom}[2]{{#1 \choose #2}} 
\newtheorem*{lemma*}{Lemma} 
\newtheorem{theorem}{Theorem} 
\newtheorem*{proposition*}{Proposition} 
\newtheorem{corollary}{Corollary} 

\newcolumntype{M}[1]{>{\centering\arraybackslash}m{#1}}
\newcolumntype{N}{@{}m{0pt}@{}}
\newcolumntype{?}{!{\vrule width 3pt}} 

\begin{document}


\title[Microreversibility and nonequilibrium response theory in magnetic fields]{Microreversibility and nonequilibrium response theory in magnetic fields}

\author{M Barbier and P Gaspard}

\address{Center for Nonlinear Phenomena and Complex Systems, Universit\'e Libre de Bruxelles (ULB), Code Postal 231, Campus Plaine, B-1050 Brussels, Belgium}

\ead{Maximilien.Barbier@ulb.ac.be,gaspard@ulb.ac.be}


\begin{abstract}
For open systems subjected to external magnetic fields, relations between the statistical cumulants of their fluctuating currents and their response coefficients are established at arbitrary orders in the deviations from equilibrium, as a consequence of microreversibility. These relations are systematically deduced from the extension of the fluctuation relation for this class of systems, and analyzed by using methods developed in [M. Barbier and P. Gaspard, J. Phys. A: Math. Theor. {\bf 51} (2018) 355001]. We unambiguously identify, among the statistical cumulants and their nonequilibrium responses, which of these quantities are independent and thus left unspecified by the fluctuation relation, i.e. by microreversibility. We also find the explicit expression of the dependent quantities in terms of the independent ones by means of coefficients of Euler polynomials.
\end{abstract}

\vspace{2pc}



\section{Introduction}

In an external magnetic field $\bm{B}$, the Hamiltonian motion of electrons and other electrically charged particles is no longer symmetric under the time-reversal transformation.  Nevertheless, this symmetry holds for the electrodynamics of the total system including the electric currents that generate the magnetic field $\bm{B}$. Since the latter currents and the magnetic field $\bm{B}$ are reversed under the transformation, the time-reversal symmetry remains between the two Hamiltonian motions  of charged particles in the magnetic fields $\bm{B}$ and $-\bm{B}$.  This symmetry called microreversibility~\cite{deGr,Cal} has fundamental consequences on the transport properties of open systems, e.g. mesoscopic electronic devices, that are driven out of thermodynamic equilibrium.

Close to equilibrium, microreversibility implies that the linear response coefficients obey the Onsager-Casimir reciprocity relations \cite{Ons31a,Ons31b,Cas45}. However, the aforementioned open systems are often driven far enough from equilibrium so that their nonlinear response properties become dominant.  Remarkably, recent advances have shown that microreversibility also entails relations between the nonlinear response coefficients and the statistical cumulants characterizing transport \cite{AG04,AG06,AG07,SU08,US09,AGM09,HPPG11,Gas13_1, Gas13_2,WF15}.  These relations, which generalize the Onsager-Casimir ones to the nonlinear regimes, can be deduced from the so-called \textit{fluctuation theorems} or \textit{fluctuation relations} (FR) and their extensions in the presence of magnetic fields \cite{AGM09,Gas13_1, Gas13_2,WF15,ECM93,ES94,GC95,Jar97,Kur98,LS99,Cro99,ZC03,Kur00,TN05,JS07,SJ08,EHM09,CHT11,LLS12}.  In the latter case, the first few relations at low orders have already been deduced \cite{SU08,US09,AGM09} and experimentally studied in Aharonov-Bohm rings \cite{NYH10,NYH11}. However, the general form of these relations and their properties have not yet been systematically investigated for an arbitrarily high order.

The purpose of the present paper is to fill this gap by extending the methods we have developed in a previous paper~\cite{BG18} to open systems in external magnetic fields.  These systems are assumed to be in contact with reservoirs of energy or particles at fixed values of temperature or chemical potential. Under such conditions, fluctuating currents flow across the open system and transfer energy or particles between the reservoirs. In the long-time limit, the statistical properties of these currents are supposed to become stationary.  They include the mean values of the currents, their diffusivities, as well as their higher statistical cumulants, as functions of the temperatures and chemical potentials of the reservoirs. This full current statistics, conventionally referred to as full counting statistics in the context of mesoscopic transport of electrons \cite{Naz}, characterizes the equilibrium and nonequilibrium steady states of the system. To consider the effect of an external magnetic field is especially relevant regarding mesoscopic transport, for instance transport through quantum dots, quantum point contacts, or Aharonov-Bohm rings connected to reservoirs of electrons. Large bias voltages between the reservoirs can be implemented without serious difficulty in related experiments (see e.g. \cite{NYH10,NYH11}), which hence induce strong nonlinear effects in the response of the system. To investigate the general nonlinear transport properties of a nonequilibrium system in a magnetic field is thus of both fundamental and practical interest.

Our starting point is the FR derived from microreversibility in the presence of an external magnetic field and expressed as a symmetry property of the generating function of the statistical cumulants of the currents \cite{SU08,AGM09,Gas13_2}. We first discuss how a set of relations satisfied by the cumulants and their nonequilibrium responses of any order can be inferred from the FR. We then unambiguously identify, among the complete set of cumulants and their responses, which of these quantities are independent and left unspecified by the FR. In addition, we explicitly write the remaining dependent quantities as linear combinations of the independent ones, the coefficients of which being related to Euler polynomials. The main outcome of our work is thus to generalize the findings of \cite{AG07,BG18}, where no magnetic field is considered, to the case of a nonzero magnetic field.

This paper begins with a concise discussion of FR in section~\ref{fluct_rel_sec}, before we introduce the statistical cumulants and their responses to the nonequilibrium constraints in section~\ref{coef_sec}. In particular, we derive from the FR a set of relations between the cumulants and their responses. We illustrate in section~\ref{rel_resp_coef_sec} how the latter relations can be used to recover the Onsager-Casimir reciprocity relations, as well as to obtain Onsager-Casimir-like relations for the second and third response coefficients. We then present our most important results in section~\ref{indep_sec}, where we perform a general analysis of the mathematical structure of the aforementioned set of relations between the cumulants and their responses. The presence of an external magnetic field is dealt with by decomposing the latter quantities into symmetric and antisymmetric parts with respect to the magnetic field. This allows to adequately reformulate some results previously obtained  in the absence of a magnetic field \cite{AG07,BG18}. Interestingly, we also derive a general identity satisfied by coefficients of Euler polynomials. Finally, we summarize our findings and give concluding remarks in section~\ref{conclusion_sec}.


\section{Fluctuation relation in an external magnetic field}\label{fluct_rel_sec}

We consider an open system coupled to $r\geqslant 2$ ideal reservoirs of energy and particles. The total system is subjected to an external static magnetic field $\bm{B}$. Its dynamics is governed by the Hamiltonian $\hat H (\bm{B})$, which satisfies the symmetry
\begin{eqnarray}
\hat\Theta \hat H (\bm{B}) \hat\Theta^{-1} = \hat H (- \bm{B})
\label{microrev_Ham}
\end{eqnarray}
under the antiunitary time-reversal operator $\hat\Theta$. The symmetry~\eref{microrev_Ham} is the general expression of microreversibility in an external magnetic field.

The reservoirs are assumed to have constant temperatures $T_j$ and chemical potentials $\mu_j$ ($j = 1,\ldots, r$). Differences in temperatures and chemical potentials drive the system out of equilibrium and generate net currents of energy and particles. We can identify these currents as flowing between a reference reservoir, say the reservoir $j=r$, and every other reservoir $j = 1,\ldots, r-1$.  In this situation, the parameters that control the nonequilibrium driving can be taken as the thermal and chemical affinities \cite{deGr,Cal,deDon,Prig} 
\begin{eqnarray}
A_{j \mathrm{E}} \equiv \frac{1}{k_{\mathrm{B}} T_r} - \frac{1}{k_{\mathrm{B}} T_j} \qquad\mbox{and}\qquad A_{j \mathrm{N}} \equiv \frac{\mu_j}{k_{\mathrm{B}} \, T_j} - \frac{\mu_r}{k_{\mathrm{B}} \, T_r}
\label{therm_chem_aff_def}
\end{eqnarray}
where $k_{\mathrm{B}}$ denotes Boltzmann's constant. At equilibrium, all reservoirs share the same temperature and chemical potential, hence making the affinities vanish together with the mean currents. For compactness, we gather in the sequel the affinities~\eref{therm_chem_aff_def} into the vector $\bm{A}$.

In order to establish the full current statistics, we use the well-known two-point quantum measurement scheme \cite{AGM09,Kur00,EHM09}. We collectively denote by $\Delta\bm{x}=\{\Delta x_k\}_{k=1}^{\chi}$ the amounts of energy and particles exchanged between the reservoirs and the reference one during the time interval $[0,t]$.  Here and in the sequel $\chi$ stands for the total number of currents.  There are as many currents as there are nonvanishing affinities that drive them, so that the vectors $\Delta\bm{x}$ and $\bm{A}$ both have the same number of components. Initially, the reservoirs are supposed to be decoupled and described by the grand-canonical density operator corresponding to their temperatures $T_j$ and chemical potentials $\mu_j$.  Before the initial time $t=0$, the energy and particle number of every reservoir are measured. The coupling is switched on at time $t=0$ and the total system follows the dynamics governed by the Hamiltonian $\hat H (\bm{B})$.  At some arbitrary time $t>0$, the coupling is switched off and the energy and particle number of every reservoir are again measured. This two-point measurement scheme allows us to obtain the transferred amounts $\Delta x_k$ of energy and particles and thus their probability density $p_t\left( \Delta\bm{x}, \bm{A} ; \bm{B} \right)$.  We note that this probability density depends on the temperatures and chemical potentials of all the reservoirs and thus on the affinities $\bm{A}$ because of the initial grand-canonical state. The generating function (GF) of the statistical moments of the fluctuating currents is then defined \cite{vanK} by the Laplace transform
\begin{eqnarray}
G_t \left( \bm{\lambda}, \bm{A} ; \bm{B} \right) \equiv \int p_t ( \Delta\bm{x}, \bm{A} ; \bm{B}) \exp( -\bm{\lambda}\cdot\Delta\bm{x}) \, d\Delta\bm{x}
\label{mom_GF_def}
\end{eqnarray}
in terms of the variables $\bm{\lambda}=\{\lambda_k\}_{k=1}^{\chi}$, which are often called the counting parameters (or counting fields). The counting parameter $\lambda_k$ is associated with the quantity $\Delta x_k$, and hence to the affinity $A_k$. Accordingly, the vectors $\bm{\lambda}$ and $\bm{A}$ both have precisely $\chi$ components.  We now consider the long-time limit $t \to \infty$ and we introduce the cumulant GF
\begin{eqnarray}
Q \left( \bm{\lambda} , \bm{A} ; \bm{B} \right) \equiv - \lim_{t \to \infty} \frac{1}{t} \, \ln G_t \left( \bm{\lambda}, \bm{A} ; \bm{B} \right)
\label{cumulant_GF_def}
\end{eqnarray}
in terms of the corresponding moment GF~\eref{mom_GF_def}.  Since this cumulant GF is defined in the long-time limit $t \to \infty$, it characterizes the full current statistics of the steady state corresponding to the affinities $\bm{A}$. In the particular case of noninteracting electrons, the cumulant GF~\eref{cumulant_GF_def} can be determined by using the scattering approach to quantum transport \cite{Gas13_1,Gas13_2,Naz,L57,L70,B86a,B86b,LL93,LS11}.

As a consequence of the symmetry~\eref{microrev_Ham} under the time-reversal transformation in the presence of an external static magnetic field $\bm{B}$, the cumulant GF~\eref{cumulant_GF_def} satisfies
\begin{eqnarray}
Q \left( \bm{\lambda} , \bm{A} ; \bm{B} \right) = Q \left( \bm{A} - \bm{\lambda} , \bm{A} ; -\bm{B} \right)
\label{fluct_rel}
\end{eqnarray}
as shown in~\cite{SU08,AGM09,Gas13_1,Gas13_2}.  This (multivariate) \textit{fluctuation relation} (FR) remains valid arbitrarily far from equilibrium and can thus be used to study transport in linear and nonlinear regimes, as is discussed here below.


\section{Statistical cumulants and response theory}\label{coef_sec}

In this section we discuss how the cumulant GF~\eref{cumulant_GF_def} is related to response theory. The latter is characterized by the coefficients of the Taylor expansion of the GF in power series of the counting parameters $\bm{\lambda}$ and the affinities $\bm{A}$. The magnetic field $\bm{B}$ is by assumption static and external to the system, and hence is treated as a fixed parameter.

First, expanding the GF in the power series
\begin{eqnarray}
Q \left( \bm{\lambda} , \bm{A} ; \bm{B} \right) = \sum_{m = 0}^{\infty} \frac{1}{m!} \, Q_{\alpha_1 \cdots \alpha_m} \left( \bm{A} ; \bm{B} \right) \lambda_{\alpha_1} \cdots \lambda_{\alpha_m}
\label{Q_exp_count_par}
\end{eqnarray}
where Einstein's summation convention for repeated indices is used, generates the statistical cumulants
\begin{eqnarray}
\fl Q_{\alpha_1 \cdots \alpha_m} \left( \bm{A} ; \bm{B} \right) \equiv \left. \frac{\partial^m Q}{\partial \lambda_{\alpha_1} \cdots \partial \lambda_{\alpha_m}} \left( \bm{\lambda}, \bm{A} ; \bm{B} \right) \right|_{\bm{\lambda} = \bm{0}} \equiv \frac{\partial^m Q}{\partial \lambda_{\alpha_1} \cdots \partial \lambda_{\alpha_m}} \left( \bm{0}, \bm{A} ; \bm{B} \right) \, .
\label{m_cumulant_def}
\end{eqnarray}
We emphasize that in both~\eref{Q_exp_count_par} and~\eref{m_cumulant_def} each subscript $\alpha_j$, $j = 1, \ldots , m$, can take any value between 1 and $\chi$ ($\chi$ being the total number of currents). Note that the definition~\eref{m_cumulant_def} can indeed be extended to the case $m=0$, the $0$th cumulant being merely identically zero as a direct consequence of the normalization condition
\begin{eqnarray}
Q \left( \bm{0} , \bm{A} ; \bm{B} \right) = 0 \, ,
\label{cumulant_GF_norm_cond}
\end{eqnarray}
which can be readily obtained from~\eref{mom_GF_def} and~\eref{cumulant_GF_def}. Substituting~\eref{cumulant_GF_norm_cond} into~\eref{fluct_rel}, we obtain the so-called `global detailed balance relation' \cite{FB08}
\begin{eqnarray}
Q \left( \bm{0} , \bm{A} ; \bm{B} \right) = Q \left( \bm{A} , \bm{A} ; - \bm{B} \right) = 0 \, ,
\label{GDBC}
\end{eqnarray}
which is weaker than the symmetry~\eref{fluct_rel} resulting from microreversibility. Indeed, we should emphasize that the condition~\eref{GDBC} is nothing but a particular case of the FR~\eref{fluct_rel}, namely for the value $\bm{\lambda}=\bm{0}$ of the counting parameters. Microreversibility on the other hand ensures that the symmetry~\eref{fluct_rel} is satisfied \textit{for any value of} $\bm{\lambda}$.

We then expand in~\eref{Q_exp_count_par} the cumulants $Q_{\alpha_1 \cdots \alpha_m} \left( \bm{A} ; \bm{B} \right)$ in power series of $\bm{A}$ to get the expansion of the GF $Q$ in power series of both the counting parameters and the affinities, and we write
\begin{eqnarray}
Q \left( \bm{\lambda} , \bm{A} ; \bm{B} \right) = \sum_{m , n = 0}^{\infty} \frac{1}{m! n!} \, Q_{\alpha_1 \cdots \alpha_m \, , \, \beta_1 \cdots \beta_n} (\bm{B}) \lambda_{\alpha_1} \cdots \lambda_{\alpha_m} A_{\beta_1} \cdots A_{\beta_n}
\label{Q_exp_count_par_and_aff}
\end{eqnarray}
where we recall that Einstein's convention is used. In~\eref{Q_exp_count_par_and_aff} we defined the quantities
\begin{eqnarray}
Q_{\alpha_1 \cdots \alpha_m \, , \, \beta_1 \cdots \beta_n} (\bm{B}) \equiv \frac{\partial^{m+n} Q}{\partial \lambda_{\alpha_1} \cdots \partial \lambda_{\alpha_m} \partial A_{\beta_1} \cdots \partial A_{\beta_n}} \left( \bm{0}, \bm{0} ; \bm{B} \right) \, ,
\label{m_cumulant_n_resp_def}
\end{eqnarray}
which are meaningful for $m,n=0$ if we prescribe that (i) for $m=0$, we have $Q_{\, , \, \beta_1 \cdots \beta_n}(\bm{B}) = 0$ because of the normalization condition~\eref{cumulant_GF_norm_cond}, while (ii) for $n=0$, we merely get the $m$th statistical cumulant, i.e. $Q_{\alpha_1 \cdots \alpha_m \, ,} (\bm{B}) = Q_{\alpha_1 \cdots \alpha_m} \left( \bm{0} ; \bm{B} \right)$. Here again, each subscript $\alpha_j$, $j = 1, \ldots , m$, and $\beta_k$, $k = 1, \ldots , n$, in~\eref{Q_exp_count_par_and_aff}-\eref{m_cumulant_n_resp_def} can take any value between 1 and $\chi$.

It is clear on its definition~\eref{m_cumulant_n_resp_def} that we have the following invariance under any permutation of the subscripts on each side of the comma:
\begin{eqnarray}
Q_{\alpha_1 \cdots \alpha_m \, , \, \beta_1 \cdots \beta_n} (\bm{B}) = Q_{\alpha_{P_m (1)} \cdots \alpha_{P_m (m)} \, , \, \beta_{P_n (1)} \cdots \beta_{P_n (n)}} (\bm{B})
\label{Q_inv_perm_indices}
\end{eqnarray}
where $P_l$ denotes a permutation of $l$ elements ($l=m,n$). Indeed, the $m$ (resp. $n$) subscripts $\alpha$ (resp. $\beta$) on the left (resp. right) of the comma denote partial derivatives of the cumulant GF with respect to counting parameters $\bm{\lambda}$ (resp. affinities $\bm{A}$), which can be performed in an arbitrary order.

The quantity $Q_{\alpha_1 \cdots \alpha_m \, , \, \beta_1 \cdots \beta_n}(\bm{B})$ is the $n$th-order response of the $m$th cumulant $Q_{\alpha_1 \cdots \alpha_m}(\bm{B})$ about the equilibrium state, corresponding to $\bm{A}=\bm{0}$ by definition of the affinities. We recall that the first cumulant $Q_{\alpha}(\bm{B})$ precisely corresponds to the mean value of the current driven by the affinity $A_{\alpha}$. The quantities $Q_{\alpha \, , \, \beta_1 \cdots \beta_n}(\bm{B})$ hence describe the response of the system to the nonequilibrium constraints and are called the \textit{response coefficients}. The linear response of the system is thus described by the first-order response coefficients $Q_{\alpha \, , \, \beta}(\bm{B})$, which satisfy the well-known Onsager-Casimir reciprocity relations \cite{Ons31a,Ons31b,Cas45}, the Green-Kubo formulae \cite{Gre52,Gre54,Kub57}, and the fluctuation-dissipation theorem \cite{CW51,Kub66}. The nonlinear response of the system is then characterized by the set of all higher-order response coefficients $Q_{\alpha \, , \, \beta_1 \cdots \beta_n}(\bm{B})$ with $n>1$. General features of the nonlinear regime can be deduced from the FR~\eref{fluct_rel}, for the latter is valid arbitrarily far from equilibrium.

Indeed, the FR~\eref{fluct_rel} can be shown \cite{AG04,AG07,SU08,AGM09,WF15,BG18} to generate Onsager-Casimir-like relations for the nonlinear response coefficients. Such relations can be conveniently obtained by expanding both sides of the FR~\eref{fluct_rel} in power series of the counting parameters $\bm{\lambda}$ and the affinities $\bm{A}$. Following \cite{BG18}, this can be for instance done by recognizing in the right-hand side of~\eref{fluct_rel} the action of the translation operator
\begin{eqnarray}
\hat T \left( \bm{A} \right) \equiv \mathrm{e}^{\bm{A} \cdot \frac{\partial}{\partial \bm{\lambda}}} \, ,
\label{transl_op_def}
\end{eqnarray}
which transforms an arbitrary function $f$ of the variable $\bm{\lambda}$ according to $\hat T \left( \bm{A} \right) f \left( \bm{\lambda} \right) = f \left( \bm{\lambda} + \bm{A} \right)$. Therefore, rewriting the right-hand side of~\eref{fluct_rel} by means of $\hat T \left( \bm{A} \right)$, expanding both sides in power series of $\bm{\lambda}$ and $\bm{A}$, and identifying on both sides the coefficients of a same power of both $\bm{\lambda}$ and $\bm{A}$, we show (following the exact same reasoning as in \cite{BG18}, which for completeness is recalled in~\ref{proof_rel_app}) that the FR yields the general relation
\begin{eqnarray}
Q_{\alpha_1 \cdots \alpha_m \, , \, \beta_1 \cdots \beta_n} (\bm{B}) = (-1)^m \sum_{j=0}^{n} Q_{\alpha_1 \cdots \alpha_m \, , \, \beta_1 \cdots \beta_n}^{(j)} (-\bm{B})
\label{gen_rel_cumul_resp}
\end{eqnarray}
for any integers $m,n \geqslant 0$, with $Q^{(0)} \equiv Q$ and
\begin{eqnarray}
Q_{\alpha_1 \cdots \alpha_m \, , \, \beta_1 \cdots \beta_n}^{(j)} (\bm{B}) \equiv \sum_{k_1 = 1}^{n} \sum_{k_{2}=1 \atop k_{2} > k_{1}}^{n} \cdots \sum_{k_{j}=1 \atop k_{j} > k_{j-1}}^{n} Q_{\alpha_1 \cdots \alpha_m \beta_{k_1} \cdots \beta_{k_j} \, , \, (\bm{\cdot})} (\bm{B})
\label{Q_j_expr}
\end{eqnarray}
for $j \geqslant 1$ and with the notation $(\bm{\cdot})$ for the set of all subscripts $\beta$ that are different of the subscripts $\beta_{k_1} , \ldots , \beta_{k_j}$ present on the left of the comma.  We notice that, for $m=0$ and any integer $n$, equation~\eref{gen_rel_cumul_resp} also results from the `global detailed balance relation'~\eref{GDBC}.

Being valid for any integers $m,n \geqslant 0$, the result~\eref{gen_rel_cumul_resp} can be adequately used to obtain Onsager-Casimir-like relations for the nonlinear response coefficients of any order. To this end, we note that, in view of the definition~\eref{Q_j_expr} of $Q^{(j)}$, the expression~\eref{gen_rel_cumul_resp} relates quantities that all possess the same total number of subscripts. Our strategy is thus to fix an arbitrary total number $\mathcal{N} = m+n$ of subscripts and to consider all the $\mathcal{N}+1$ possible relations~\eref{gen_rel_cumul_resp} that correspond to this given $\mathcal{N}$ for all the possible values of the indices $m$ and $n$. We illustrate this procedure in section~\ref{rel_resp_coef_sec}, where we consider simple values of $\mathcal{N}$, namely $\mathcal{N}=2,3,4$, to explicitly derive the relations satisfied by the first-, second- and third-order response coefficients. Section~\ref{indep_sec} is then devoted to a general analysis of the mathematical structure of the relations~\eref{gen_rel_cumul_resp}. Our aim is to unambiguously identify the independent quantities that are left unspecified by the FR, and thus to generalize the results obtained in \cite{AG07,BG18} in the absence of a magnetic field to the case of a nonzero magnetic field.


\section{Consequences of microreversibility at low orders}\label{rel_resp_coef_sec}

Here we discuss in more details how the general result~\eref{gen_rel_cumul_resp} generates relations of the Onsager-Casimir type satisfied by the response coefficients. We do this by investigating all possible relations obtained from~\eref{gen_rel_cumul_resp} for a fixed total number $\mathcal{N} = m+n$ of subscripts and all possible values of $m$ and $n$. We illustrate the procedure on the response coefficients of first (i.e. for $\mathcal{N}=2$), second ($\mathcal{N}=3$) and third ($\mathcal{N}=4$) orders in subsections~\ref{first_order_subsec},~\ref{second_order_subsec} and~\ref{third_order_subsec}, respectively. We also briefly present in subsection~\ref{parity_cumulants_sec} a parity property, with respect to the magnetic field $\bm{B}$, satisfied by the statistical cumulants as a direct consequence of the general result~\eref{gen_rel_cumul_resp}.

\begin{table}[ht]
\centering
\caption{Comparison between the quantities $Q_{\alpha_1 \cdots \alpha_m \, , \, \beta_1 \cdots \beta_n} (\bm{B})$ defined by~\eref{m_cumulant_n_resp_def} and notations used in \cite{AGM09} (section 5) and \cite{Gas13_2} (section V) in the study of the response coefficients of first, second and third orders, which correspond to the cases $\mathcal{N}=2,3$ and 4, respectively. The case $\mathcal{N}=1$ corresponds to the first cumulants, i.e. the mean currents.}
\vskip 0.2 cm
\begin{tabular}{ |M{0.5cm}|M{2.5cm}|M{3cm} M{2cm}| }
\hline
$\mathcal{N}$ & Present notations & Previous notations & References \\ \hline \hline
\multirow{1}{*}{1} & $Q_{\alpha} \left( \bm{A} ; \bm{B} \right)$ & $J_{\alpha} \left( \bm{A} ; \bm{B} \right)$ & \cite{AGM09,Gas13_2} \\ \hline
\multirow{2}{*}{2} & $Q_{\alpha \beta} \left( \bm{A} ; \bm{B} \right)$ & $-2 D_{\alpha \beta} \left( \bm{A} ; \bm{B} \right)$ & \cite{Gas13_2} \\
 & $Q_{\alpha , \beta} (\bm{B})$ & $L_{\alpha , \beta} (\bm{B})$ & \cite{AGM09,Gas13_2} \\ \hline
\multirow{3}{*}{3} & $Q_{\alpha \beta \gamma} \left( \bm{A} ; \bm{B} \right)$ & $C_{\alpha \beta \gamma} \left( \bm{A} ; \bm{B} \right)$ & \cite{Gas13_2} \\
 & $Q_{\alpha \beta , \gamma} (\bm{B})$ & $-R_{\alpha \beta , \gamma} (\bm{B})$ & \cite{AGM09} \\
 & $Q_{\alpha , \beta \gamma} (\bm{B})$ & $M_{\alpha , \beta \gamma} (\bm{B})$ & \cite{AGM09,Gas13_2} \\ \hline
\multirow{4}{*}{4} & $Q_{\alpha \beta \gamma \delta} \left( \bm{A} ; \bm{B} \right)$ & $-2 B_{\alpha \beta \gamma \delta} \left( \bm{A} ; \bm{B} \right)$ & \cite{Gas13_2} \\
 & $Q_{\alpha \beta \gamma , \delta} (\bm{B})$ & $S_{\alpha \beta \gamma , \delta} (\bm{B})$ & \cite{AGM09} \\
 & $Q_{\alpha \beta , \gamma \delta} (\bm{B})$ & $-T_{\alpha \beta , \gamma \delta} (\bm{B})$ & \cite{AGM09} \\
 & $Q_{\alpha , \beta \gamma \delta} (\bm{B})$ & $N_{\alpha , \beta \gamma \delta} (\bm{B})$ & \cite{AGM09,Gas13_2} \\
\hline
\end{tabular}
\label{table1}
\end{table}

Such an investigation has already been performed e.g. in \cite{AGM09} and \cite{Gas13_2} for $\mathcal{N}=2$ and~3, while the case $\mathcal{N}=4$ is briefly treated in \cite{AGM09} but not discussed in details. To simplify a comparison between the relations obtained in the sequel and corresponding results of \cite{AGM09} and \cite{Gas13_2}, we give in table~\ref{table1} the connection between the relevant quantities $Q_{\alpha_1 \cdots \alpha_m \, , \, \beta_1 \cdots \beta_n} (\bm{B})$ and the corresponding notations used in \cite{AGM09} and \cite{Gas13_2}.


\subsection{First order}\label{first_order_subsec}

Here we consider the three different relations obtained from~\eref{gen_rel_cumul_resp} for a fixed $\mathcal{N} = m+n = 2$. Taking first $m=2$ and $n=0$ and setting $\alpha_1 , \alpha_2 = \alpha , \beta$ yields
\begin{eqnarray}
Q_{\alpha \beta} \left( \bm{0} ; \bm{B} \right) = Q_{\alpha \beta} \left( \bm{0} ; - \bm{B} \right) \, .
\label{sec_cumu_even_mag_field}
\end{eqnarray}
This shows that the second cumulants at equilibrium (where $\bm{A} = \bm{0}$) are even with respect to the magnetic field $\bm{B}$.

We then consider~\eref{gen_rel_cumul_resp} for $m=1$ and $n=1$. Setting $\alpha_1 , \beta_1 = \alpha , \beta$ and using the symmetry~\eref{sec_cumu_even_mag_field}, the first response coefficient $Q_{\alpha , \beta}$ can be expressed only in terms of the second cumulant $Q_{\alpha \beta}$ through
\begin{eqnarray}
Q_{\alpha , \beta} (\bm{B}) + Q_{\alpha , \beta} (-\bm{B}) = - Q_{\alpha \beta} \left( \bm{0} ; \bm{B} \right) \, .
\label{first_resp_coef_from_sec_cumu}
\end{eqnarray}

Finally, we take $m=0$ and $n=2$ into~\eref{gen_rel_cumul_resp}. Setting $\beta_1 , \beta_2 = \alpha , \beta$ and reversing the magnetic field ($\bm{B} \to - \bm{B}$) yields
\begin{eqnarray}
Q_{\alpha , \beta} (\bm{B}) + Q_{\beta,\alpha} \left(\bm{B} \right) = - Q_{\alpha \beta} \left( \bm{0} ; \bm{B} \right) \, ,
\label{m=0_n=2}
\end{eqnarray}
which also results from the `global detailed balance relation'~\eref{GDBC}.  The difference between~\eref{first_resp_coef_from_sec_cumu} and~\eref{m=0_n=2} gives, after reversing the magnetic field,
\begin{eqnarray}
Q_{\alpha , \beta} (\bm{B}) = Q_{\beta , \alpha} (-\bm{B}) \, ,
\label{Onsager_Casimir_rel}
\end{eqnarray}
which are precisely the Onsager-Casimir reciprocity relations \cite{Ons31a,Ons31b,Cas45} in the presence of a magnetic field.

We now illustrate how the FR generates relations of the Onsager-Casimir type for the nonlinear response coefficients. This is done for the response coefficients of second and third orders in subsections~\ref{second_order_subsec} and~\ref{third_order_subsec}, respectively.


\subsection{Second order}\label{second_order_subsec}

We now consider the four relations obtained from~\eref{gen_rel_cumul_resp} for a fixed $\mathcal{N} = m+n = 3$. We first take $m=3$ and $n=0$ and set $\alpha_1 , \alpha_2 , \alpha_3 = \alpha , \beta , \gamma$ to get
\begin{eqnarray}
Q_{\alpha \beta \gamma} \left( \bm{0} ; \bm{B} \right) = - Q_{\alpha \beta \gamma} \left( \bm{0} ; - \bm{B} \right) \, .
\label{third_cumu_odd_mag_field}
\end{eqnarray}
This shows that the third cumulants at equilibrium are odd with respect to the magnetic field $\bm{B}$. It is thus clear on~\eref{third_cumu_odd_mag_field} that in the absence of a magnetic field, $\bm{B} = \bm{0}$, the third cumulants vanish at equilibrium, i.e.
\begin{eqnarray}
Q_{\alpha \beta \gamma} \left( \bm{0} ; \bm{0} \right) = 0 \, .
\label{zero_third_cumu_without_mag_field}
\end{eqnarray}

Then we consider the relation~\eref{gen_rel_cumul_resp} for $m=2$ and $n=1$. Setting $\alpha_1 , \alpha_2 , \beta_1 = \alpha , \beta , \gamma$ and using the symmetry~\eref{third_cumu_odd_mag_field} yields
\begin{eqnarray}
Q_{\alpha \beta , \gamma} (\bm{B}) - Q_{\alpha \beta , \gamma} (-\bm{B}) = - Q_{\alpha \beta \gamma} \left( \bm{0} ; \bm{B} \right) \, .
\label{resp_sec_cumu_from_third_cumu}
\end{eqnarray}
The third cumulant $Q_{\alpha \beta \gamma}$ can thus be expressed in terms of the first response $Q_{\alpha \beta , \gamma}$ of second cumulants only.

We now take $m=1$ and $n=2$ into~\eref{gen_rel_cumul_resp}, set $\alpha_1 , \beta_1 , \beta_2 = \alpha , \beta , \gamma$ and use the result~\eref{resp_sec_cumu_from_third_cumu} to get
\begin{eqnarray}
Q_{\alpha , \beta \gamma} (\bm{B}) + Q_{\alpha , \beta \gamma} (-\bm{B}) = - Q_{\alpha \beta , \gamma} (\bm{B}) - Q_{\alpha \gamma , \beta} (-\bm{B}) \, ,
\label{sec_resp_coef_single_indices_resp_sec_cumu}
\end{eqnarray}
which relates the second response coefficient $Q_{\alpha , \beta \gamma}$ to first responses of second cumulants.

Finally, we consider~\eref{gen_rel_cumul_resp} for $m=0$ and $n=3$. Setting $\beta_1 , \beta_2 , \beta_3 = \alpha , \beta , \gamma$ and reversing the magnetic field yields
\begin{eqnarray}
\fl Q_{\alpha , \beta \gamma} (\bm{B}) + Q_{\beta , \gamma \alpha} (\bm{B}) + Q_{\gamma , \alpha \beta} (\bm{B}) \nonumber\\[0.15cm]
= - Q_{\alpha \beta , \gamma} (\bm{B}) - Q_{\gamma \alpha , \beta} (\bm{B}) - Q_{\beta \gamma , \alpha } (\bm{B}) - Q_{\alpha \beta \gamma} \left( \bm{0} ; \bm{B} \right) \, ,
\label{sec_resp_coef_cyclic_indices}
\end{eqnarray}
which hence expresses a sum of second response coefficients of all cyclic permutations of the indices $\alpha$, $\beta$ and $\gamma$ in terms of the third cumulant $Q_{\alpha \beta \gamma}$ and first responses of second cumulants. Equation~\eref{sec_resp_coef_cyclic_indices} is also a consequence of the `global detailed balance relation'~\eref{GDBC}.

We now take the difference of the result~\eref{sec_resp_coef_cyclic_indices} for $\bm{B}$ and $- \bm{B}$ and get, in view of the results~\eref{third_cumu_odd_mag_field} and~\eref{resp_sec_cumu_from_third_cumu},
\begin{eqnarray}
\fl Q_{\alpha \beta \gamma} \left( \bm{0} ; \bm{B} \right) = Q_{\alpha , \beta \gamma} (\bm{B}) + Q_{\beta , \gamma \alpha} (\bm{B}) + Q_{\gamma , \alpha \beta} (\bm{B}) \nonumber\\[0.15cm]
- Q_{\alpha , \beta \gamma} (-\bm{B}) - Q_{\beta , \gamma \alpha} (-\bm{B}) - Q_{\gamma , \alpha \beta} (-\bm{B}) \, .
\label{third_cumu_sec_resp_coef}
\end{eqnarray}
This result clearly shows that the third cumulant $Q_{\alpha \beta \gamma}$ is fully determined by the second response coefficients. Note that in the absence of a magnetic field we immediately recover from~\eref{third_cumu_sec_resp_coef} the previously-established result~\eref{zero_third_cumu_without_mag_field}.

We conclude this section by analyzing the special case where all three indices are identical. Setting $\alpha = \beta = \gamma$ into~\eref{third_cumu_sec_resp_coef}, we immediately get
\begin{eqnarray}
Q_{\alpha \alpha \alpha} \left( \bm{0} ; \bm{B} \right) = 3 \left[ Q_{\alpha , \alpha \alpha} (\bm{B}) - Q_{\alpha , \alpha \alpha} (-\bm{B}) \right]
\label{third_cumu_sec_resp_coef_same_indices}
\end{eqnarray}
while~\eref{sec_resp_coef_cyclic_indices} reads, for $\alpha = \beta = \gamma$,
\begin{eqnarray}
Q_{\alpha , \alpha \alpha} (\bm{B}) = - Q_{\alpha \alpha , \alpha} (\bm{B}) - \frac{1}{3} Q_{\alpha \alpha \alpha} \left( \bm{0} ; \bm{B} \right) \, .
\label{sec_resp_coef_identical_indices_resp_sec_cumu}
\end{eqnarray}
It is now worth noting that combining the two results~\eref{third_cumu_sec_resp_coef_same_indices} and~\eref{sec_resp_coef_identical_indices_resp_sec_cumu} readily yields
\begin{eqnarray}
Q_{\alpha \alpha , \alpha} (\bm{B}) = Q_{\alpha , \alpha \alpha} (-\bm{B}) - 2 Q_{\alpha , \alpha \alpha} (\bm{B}) \, .
\label{resp_sec_cumu_sec_resp_coef_same_indices}
\end{eqnarray}
Therefore, the first response of any second cumulant of two same indices $\alpha$ with respect to the affinity $A$ of the same index $\alpha$ turns out to be fully specified by the second response coefficient of indices $\alpha$. In the absence of a magnetic field, $\bm{B} = \bm{0}$, the relation~\eref{resp_sec_cumu_sec_resp_coef_same_indices} merely reads
\begin{eqnarray}
Q_{\alpha \alpha , \alpha} \left( \bm{0} \right) = - Q_{\alpha , \alpha \alpha} \left( \bm{0} \right) \, .
\label{resp_sec_cumu_sec_resp_coef_same_indices_zero_mag_field}
\end{eqnarray}

We now analyze the consequences of the result~\eref{gen_rel_cumul_resp} on the third-order response coefficients.


\subsection{Third order}\label{third_order_subsec}

Here we consider the five relations obtained from~\eref{gen_rel_cumul_resp} for a fixed $\mathcal{N} = m+n = 4$. Taking first $m=4$ and $n=0$ into~\eref{gen_rel_cumul_resp} and setting $\alpha_1 , \alpha_2 , \alpha_3, \alpha_4 = \alpha , \beta , \gamma , \delta$ yields
\begin{eqnarray}
Q_{\alpha \beta \gamma \delta} \left( \bm{0} ; \bm{B} \right) = Q_{\alpha \beta \gamma \delta} \left( \bm{0} ; - \bm{B} \right) \, .
\label{fourth_cumu_even_mag_field}
\end{eqnarray}
As is the case for the second cumulants, the fourth cumulants at equilibrium are thus even with respect to the magnetic field $\bm{B}$.

We now consider the relation~\eref{gen_rel_cumul_resp} for $m=3$ and $n=1$. We set $\alpha_1 , \alpha_2 , \alpha_3, \beta_1 = \alpha , \beta , \gamma , \delta$ and use the symmetry~\eref{fourth_cumu_even_mag_field} to obtain
\begin{eqnarray}
Q_{\alpha \beta \gamma , \delta} (\bm{B}) + Q_{\alpha \beta \gamma , \delta} (-\bm{B}) = - Q_{\alpha \beta \gamma \delta} \left( \bm{0} ; \bm{B} \right) \, .
\label{resp_third_cumu_fourth_cumu}
\end{eqnarray}
This result has a structure similar to~\eref{resp_sec_cumu_from_third_cumu}, for it shows that the fourth cumulant $Q_{\alpha \beta \gamma \delta}$ can be expressed in terms of the first response $Q_{\alpha \beta \gamma , \delta}$ of third cumulants only.

Then we consider the relation~\eref{gen_rel_cumul_resp} for $m=2$ and $n=2$. Setting now $\alpha_1 , \alpha_2 , \beta_1, \beta_2 = \alpha , \beta , \gamma , \delta$ and using again~\eref{fourth_cumu_even_mag_field} as well as the result~\eref{resp_third_cumu_fourth_cumu}, we get
\begin{eqnarray}
Q_{\alpha \beta , \gamma \delta} (\bm{B}) - Q_{\alpha \beta , \gamma \delta} (-\bm{B}) = Q_{\alpha \beta \delta , \gamma} (-\bm{B}) - Q_{\alpha \beta \gamma , \delta} (\bm{B}) \, .
\label{sec_resp_sec_cumu_resp_third_cumu}
\end{eqnarray}
It is worth noting that an immediate consequence of the latter is the symmetry
\begin{eqnarray}
Q_{\alpha \beta \gamma , \delta} (\bm{B}) + Q_{\alpha \beta \gamma , \delta} (-\bm{B}) = Q_{\alpha \beta \delta , \gamma} (\bm{B}) + Q_{\alpha \beta \delta , \gamma} (-\bm{B})
\label{symmetry_resp_third_cumu}
\end{eqnarray}
satisfied by the first responses of third cumulants. This property can be alternatively derived from~\eref{resp_third_cumu_fourth_cumu}. Indeed, it is a direct consequence of the symmetry of $Q_{\alpha \beta \gamma \delta}$ with respect to any permutation of its indices.

The next relations are obtained from~\eref{gen_rel_cumul_resp} for $m=1$ and $n=3$. We set $\alpha_1 , \beta_1 , \beta_2, \beta_3 = \alpha , \beta , \gamma , \delta$ and reverse the magnetic field to obtain
\begin{eqnarray}
\fl Q_{\alpha , \beta \gamma \delta} (\bm{B}) + Q_{\alpha , \beta \gamma \delta} (-\bm{B}) = - Q_{\alpha \beta , \gamma \delta} (\bm{B}) - Q_{\alpha \gamma , \beta \delta} (\bm{B}) - Q_{\alpha \delta , \beta \gamma} (\bm{B}) \nonumber\\[0.15cm]
- Q_{\alpha \beta \gamma , \delta} (\bm{B}) - Q_{\delta \alpha \beta , \gamma} (\bm{B}) - Q_{\gamma \delta \alpha , \beta} (\bm{B}) - Q_{\alpha \beta \gamma \delta} \left( \bm{0} ; \bm{B} \right) \, .
\label{third_resp_coef_identical_indices}
\end{eqnarray}

Finally, we consider the relation~\eref{gen_rel_cumul_resp} for $m=0$ and $n=4$. Setting $\beta_1 , \beta_2 , \beta_3, \beta_4 = \alpha , \beta , \gamma , \delta$ and reversing the magnetic field yields
\begin{eqnarray}
\fl Q_{\alpha , \beta \gamma \delta} (\bm{B}) + Q_{\beta , \gamma \delta \alpha} (\bm{B}) + Q_{\gamma , \delta \alpha \beta} (\bm{B}) + Q_{\delta , \alpha \beta \gamma} (\bm{B}) \nonumber\\[0.15cm]
\fl = - Q_{\alpha \beta , \gamma \delta} (\bm{B}) - Q_{\alpha \gamma , \beta \delta} (\bm{B}) - Q_{\alpha \delta , \beta \gamma} (\bm{B}) - Q_{\beta \gamma , \alpha \delta} (\bm{B}) - Q_{\beta \delta , \alpha \gamma} (\bm{B}) - Q_{\gamma \delta , \alpha \beta} (\bm{B}) \nonumber\\[0.15cm]
\fl - Q_{\alpha \beta \gamma , \delta} (\bm{B}) - Q_{\delta \alpha \beta , \gamma} (\bm{B}) - Q_{\gamma \delta \alpha , \beta} (\bm{B}) - Q_{\beta \gamma \delta , \alpha} (\bm{B}) - Q_{\alpha \beta \gamma \delta} \left( \bm{0} ; \bm{B} \right) \, ,
\label{third_resp_coef_cyclic_indices}
\end{eqnarray}
which can also be deduced from the `global detailed balance relation'~\eref{GDBC}.

We now note that the two results~\eref{third_resp_coef_identical_indices} and~\eref{third_resp_coef_cyclic_indices} can, in view of the relation~\eref{resp_third_cumu_fourth_cumu}, be rewritten as
\begin{eqnarray}
\fl Q_{\alpha , \beta \gamma \delta} (\bm{B}) + Q_{\alpha , \beta \gamma \delta} (-\bm{B}) = - Q_{\alpha \beta , \gamma \delta} (\bm{B}) - Q_{\alpha \gamma , \beta \delta} (\bm{B}) - Q_{\alpha \delta , \beta \gamma} (\bm{B}) \nonumber\\[0.15cm]
+ Q_{\alpha \beta \gamma , \delta} (-\bm{B}) - Q_{\delta \alpha \beta , \gamma} (\bm{B}) - Q_{\gamma \delta \alpha , \beta} (\bm{B})
\label{third_resp_coef_identical_indices_no_cumu}
\end{eqnarray}
and
\begin{eqnarray}
\fl Q_{\alpha , \beta \gamma \delta} (\bm{B}) + Q_{\beta , \gamma \delta \alpha} (\bm{B}) + Q_{\gamma , \delta \alpha \beta} (\bm{B}) + Q_{\delta , \alpha \beta \gamma} (\bm{B}) \nonumber\\[0.15cm]
\fl = - Q_{\alpha \beta , \gamma \delta} (\bm{B}) - Q_{\alpha \gamma , \beta \delta} (\bm{B}) - Q_{\alpha \delta , \beta \gamma} (\bm{B}) - Q_{\beta \gamma , \alpha \delta} (\bm{B}) - Q_{\beta \delta , \alpha \gamma} (\bm{B}) - Q_{\gamma \delta , \alpha \beta} (\bm{B}) \nonumber\\[0.15cm]
+ Q_{\alpha \beta \gamma , \delta} (-\bm{B}) - Q_{\delta \alpha \beta , \gamma} (\bm{B}) - Q_{\gamma \delta \alpha , \beta} (\bm{B}) - Q_{\beta \gamma \delta , \alpha} (\bm{B}) \, .
\label{third_resp_coef_cyclic_indices_no_cumu}
\end{eqnarray}
The results~\eref{third_resp_coef_identical_indices_no_cumu} and~\eref{third_resp_coef_cyclic_indices_no_cumu} show that the third response coefficients can be expressed in terms of second and first responses of second and third cumulants, respectively.

We now take the difference of the result~\eref{third_resp_coef_cyclic_indices_no_cumu} for $\bm{B}$ and $- \bm{B}$ and get, in view of~\eref{sec_resp_sec_cumu_resp_third_cumu},
\begin{eqnarray}
\fl Q_{\alpha , \beta \gamma \delta} (\bm{B}) + Q_{\beta , \gamma \delta \alpha} (\bm{B}) + Q_{\gamma , \delta \alpha \beta} (\bm{B}) + Q_{\delta , \alpha \beta \gamma} (\bm{B}) \nonumber\\[0.15cm]
- Q_{\alpha , \beta \gamma \delta} (-\bm{B}) - Q_{\beta , \gamma \delta \alpha} (-\bm{B}) - Q_{\gamma , \delta \alpha \beta} (-\bm{B}) - Q_{\delta , \alpha \beta \gamma} (-\bm{B}) \nonumber\\[0.15cm]
= - Q_{\alpha \delta , \beta \gamma} (\bm{B}) + Q_{\alpha \delta , \beta \gamma} (-\bm{B}) - Q_{\beta \gamma , \alpha \delta} (\bm{B}) + Q_{\beta \gamma , \alpha \delta} (-\bm{B}) \nonumber\\[0.15cm]
= Q_{\gamma \delta \alpha , \beta} (\bm{B}) - Q_{\delta \alpha \beta , \gamma} (-\bm{B}) + Q_{\beta \gamma \delta , \alpha} (\bm{B}) - Q_{\alpha \beta \gamma , \delta} (-\bm{B}) \, .
\label{difference_all_third_resp_coef_rel}
\end{eqnarray}
We can readily see on~\eref{difference_all_third_resp_coef_rel} that a particular combination of third response coefficients can be fully expressed either in terms of second responses of second cumulants, or in terms of first responses of third cumulants. Furthermore, we can derive a finer result that relates merely four response coefficients (as compared to the eight involved in~\eref{difference_all_third_resp_coef_rel}) to two responses of second cumulants.

To show this, we first combine the result~\eref{third_resp_coef_identical_indices} for (i) $\alpha$ as the first index in the left-hand side, and magnetic field $\bm{B}$, (ii) $\beta$ as the first index, and magnetic field $- \bm{B}$, (iii) $\gamma$ as the first index, and magnetic field $\bm{B}$, and (iv) $\delta$ as the first index, and magnetic field $- \bm{B}$. Adding (i) and (iv) and then subtracting (ii) and (iii), we get, using in addition the results~\eref{fourth_cumu_even_mag_field} and~\eref{sec_resp_sec_cumu_resp_third_cumu}-\eref{symmetry_resp_third_cumu},
\begin{eqnarray}
\fl Q_{\alpha , \beta \gamma \delta} (\bm{B}) + Q_{\alpha , \beta \gamma \delta} (-\bm{B}) - Q_{\beta , \gamma \delta \alpha} (\bm{B}) - Q_{\beta , \gamma \delta \alpha} (-\bm{B}) \nonumber\\[0.15cm]
- Q_{\gamma , \delta \alpha \beta} (\bm{B}) - Q_{\gamma , \delta \alpha \beta} (-\bm{B}) + Q_{\delta , \alpha \beta \gamma} (\bm{B}) + Q_{\delta , \alpha \beta \gamma} (-\bm{B}) \nonumber\\[0.15cm]
= - Q_{\alpha \delta , \beta \gamma} (\bm{B}) - Q_{\alpha \delta , \beta \gamma} (-\bm{B}) + Q_{\beta \gamma , \alpha \delta} (\bm{B}) + Q_{\beta \gamma , \alpha \delta} (-\bm{B}) \, .
\label{combination_all_third_resp_coef_rel}
\end{eqnarray}
Besides, the sum of~\eref{difference_all_third_resp_coef_rel} and~\eref{combination_all_third_resp_coef_rel} divided by two gives
\begin{eqnarray}
\fl Q_{\alpha , \beta \gamma \delta} (\bm{B}) - Q_{\beta , \gamma \delta \alpha} (-\bm{B}) - Q_{\gamma , \delta \alpha \beta} (-\bm{B}) + Q_{\delta , \alpha \beta \gamma} (\bm{B}) = Q_{\beta \gamma , \alpha \delta} (-\bm{B}) - Q_{\alpha \delta , \beta \gamma} (\bm{B}) \, . 
\label{third_resp_coef_from_sec_resp_sec_cumu}
\end{eqnarray}
Therefore, we can relate a combination of four third response coefficients (namely, two of them are added with a same magnetic field, while the other two are subtracted with the opposite magnetic field) of all cyclic permutations of the indices $\alpha$, $\beta$, $\gamma$ and $\delta$ to a mere difference of two second responses of second cumulants. It can be readily checked that taking the difference of~\eref{third_resp_coef_from_sec_resp_sec_cumu} for $\bm{B}$ and $- \bm{B}$ gives back the relation~\eref{difference_all_third_resp_coef_rel}.

It is also worth giving another result, which is a direct consequence of the relation~\eref{sec_resp_sec_cumu_resp_third_cumu}. Combining the latter for (i) $\alpha , \beta$ as the first two indices in the left-hand side, and magnetic field $\bm{B}$, (ii) $\alpha , \gamma$ as the first two indices, and magnetic field $\bm{B}$, and (iii) $\alpha , \delta$ as the first two indices, and magnetic field $\bm{B}$, adding (i) and (iii) and then subtracting (ii), we get
\begin{eqnarray}
\fl - Q_{\alpha \beta , \gamma \delta} (\bm{B}) + Q_{\alpha \beta , \gamma \delta} (-\bm{B}) + Q_{\alpha \gamma , \beta \delta} (\bm{B}) - Q_{\alpha \gamma , \beta \delta} (-\bm{B}) - Q_{\alpha \delta , \beta \gamma} (\bm{B}) + Q_{\alpha \delta , \beta \gamma} (-\bm{B}) \nonumber\\[0.15cm]
= Q_{\delta \alpha \beta , \gamma} (\bm{B}) - Q_{\delta \alpha \beta , \gamma} (-\bm{B}) \, .
\label{sec_resp_sec_cumu_resp_third_cumu_alternative_rel}
\end{eqnarray}
Nonetheless, because of the minus sign in the right-hand side of~\eref{sec_resp_sec_cumu_resp_third_cumu_alternative_rel}, this result can not be combined with~\eref{resp_third_cumu_fourth_cumu} to express the fourth cumulants in terms of second responses of second cumulants only. Hence~\eref{resp_third_cumu_fourth_cumu} is the only result available that relates the fourth cumulants to a single kind of quantity, namely first responses of third cumulants. In particular, and contrary to the situation at the previous orders (see the relations~\eref{first_resp_coef_from_sec_cumu} and~\eref{third_cumu_sec_resp_coef}), the fourth cumulant $Q_{\alpha \beta \gamma \delta}$ can not be fully determined by the third response coefficients.

We conclude this section with a short discussion of the consequences of the general result~\eref{gen_rel_cumul_resp} on the parity of the cumulants of any order with respect to the magnetic field $\bm{B}$.


\subsection{Parity of the cumulants}\label{parity_cumulants_sec}

Here we consider the relation~\eref{gen_rel_cumul_resp} for an arbitrary $m \geqslant 0$, but for $n=0$. This readily yields
\begin{eqnarray}
Q_{\alpha_1 \cdots \alpha_m} \left( \bm{0} ; \bm{B} \right) = (-1)^m Q_{\alpha_1 \cdots \alpha_m} \left( \bm{0} ; - \bm{B} \right) \, ,
\label{parity_m_cumu}
\end{eqnarray}
which clearly shows that the statistical cumulants at equilibrium (i.e. for $\bm{A} = \bm{0}$) have a simple behavior with respect to the magnetic field $\bm{B}$. Indeed, the cumulants of even order are thus \textit{even} with respect to $\bm{B}$, while the cumulants of odd order are \textit{odd} with respect to $\bm{B}$. This means that in the absence of any magnetic field, i.e. $\bm{B} = \bm{0}$, all the cumulants of odd order hence vanish,
\begin{eqnarray}
Q_{\alpha_1 \cdots \alpha_{2k+1}} \left( \bm{0} ; \bm{0} \right) = 0
\label{odd_cumu_zero_mag_field}
\end{eqnarray}
for any $k \geqslant 0$.

We illustrated in this section how relations of the Onsager-Casimir type for the first, second and third response coefficients can be derived from the general result~\eref{gen_rel_cumul_resp}. This was done by considering all relations obtained from~\eref{gen_rel_cumul_resp} for simple, fixed values of the total number $\mathcal{N}=m+n$ of subscripts, namely $\mathcal{N}=2,3,4$. We now investigate the general mathematical structure of~\eref{gen_rel_cumul_resp} for an arbitrary $\mathcal{N} \geqslant 1$.


\section{Independent quantities}\label{indep_sec}

Our aim in this section is to unambiguously identify the independent quantities that are left unspecified by the fluctuation relation (FR)~\eref{fluct_rel}, i.e. by the general relations~\eref{gen_rel_cumul_resp}. This generalizes the study of \cite{AG07,BG18}, where no magnetic field is considered, to the case of a nonzero magnetic field. The first step is to rewrite~\eref{gen_rel_cumul_resp} so as to obtain relations similar to the case of a zero magnetic field.

We do this by splitting the quantities $Q_{\alpha_1 \cdots \alpha_m \, , \, \beta_1 \cdots \beta_n} (\bm{B})$ into symmetric and antisymmetric parts with respect to the magnetic field $\bm{B}$. Given an arbitrary function $f$ of $\bm{B}$, its symmetric and antisymmetric parts $f^{\mathrm{S}}$ and $f^{\mathrm{A}}$, respectively, are defined by
\begin{eqnarray}
f^{\mathrm{S},\mathrm{A}}(\bm{B}) \equiv \frac{1}{2} \left[ f( \bm{B}) \pm f( -\bm{B}) \right] \, .
\label{sym_antisym_gen_def}
\end{eqnarray}
The function $f$ is then uniquely specified by its symmetric and antisymmetric parts and we have
\begin{eqnarray}
f( \pm \bm{B}) = f^{\mathrm{S}}(\bm{B}) \pm f^{\mathrm{A}}(\bm{B}) \, .
\label{fct_from_sym_antisym}
\end{eqnarray}
We now infer from~\eref{gen_rel_cumul_resp} the relations satisfied by the symmetric and antisymmetric parts of the quantities $Q_{\alpha_1 \cdots \alpha_m \, , \, \beta_1 \cdots \beta_n} (\bm{B})$. To this end, it proves useful to introduce the notations
\begin{eqnarray}
\bm{\alpha}_{m} \equiv \alpha_1 \cdots \alpha_{m} \qquad\mbox{and}\qquad \bm{\beta}_{n} \equiv \beta_1 \cdots \beta_{n} \, .
\label{alpha_beta_def}
\end{eqnarray}

We first write~\eref{gen_rel_cumul_resp} for $\bm{B}$ and $-\bm{B}$, and add the two resulting relations. This produces, in view of~\eref{sym_antisym_gen_def}, the set of relations satisfied by the symmetric parts $Q^{\mathrm{S}}$, namely
\begin{eqnarray}
Q_{\bm{\alpha}_{m} \, , \, \bm{\beta}_{n}}^{\mathrm{S}} (\bm{B}) = (-1)^m \sum_{j=0}^{n} Q_{\bm{\alpha}_{m} \, , \, \bm{\beta}_{n}}^{(j) \, \mathrm{S}} (\bm{B})
\label{gen_rel_sym_parts}
\end{eqnarray}
where $Q^{(j) \, \mathrm{S}}$ denotes the symmetric part of the quantity $Q^{(j)}$ defined by~\eref{Q_j_expr}. Then, writing again~\eref{gen_rel_cumul_resp} for $\bm{B}$ and $-\bm{B}$ but subtracting the two yields the set of relations satisfied by the antisymmetric parts $Q^{\mathrm{A}}$, i.e.
\begin{eqnarray}
Q_{\bm{\alpha}_{m} \, , \, \bm{\beta}_{n}}^{\mathrm{A}} (\bm{B}) = (-1)^{m+1} \sum_{j=0}^{n} Q_{\bm{\alpha}_{m} \, , \, \bm{\beta}_{n}}^{(j) \, \mathrm{A}} (\bm{B})
\label{gen_rel_antisym_parts}
\end{eqnarray}
where $Q^{(j) \, \mathrm{A}}$ now denotes the antisymmetric part of $Q^{(j)}$.

Therefore, the FR~\eref{fluct_rel} can be seen to generate, rather than the single set of relations~\eref{gen_rel_cumul_resp}, the two distinct sets~\eref{gen_rel_sym_parts} and~\eref{gen_rel_antisym_parts} for symmetric and antisymmetric parts, respectively. The latter have the advantage of relating quantities evaluated at a same magnetic field, and not a reversed one as in~\eref{gen_rel_cumul_resp}. We hence note that the relations~\eref{gen_rel_sym_parts} and~\eref{gen_rel_antisym_parts} have exactly the same mathematical structure whether $\bm{B}$ is zero or nonzero.

We now discuss how results obtained in the absence of a magnetic field \cite{AG07,BG18} can be generalized to the relations~\eref{gen_rel_sym_parts} and~\eref{gen_rel_antisym_parts}.  Our strategy is to separately investigate all possible relations obtained from~\eref{gen_rel_sym_parts} and~\eref{gen_rel_antisym_parts} for a fixed total number $\mathcal{N} = m+n$ of subscripts and all possible values of $m$ and $n$. We first note that the relations~\eref{gen_rel_sym_parts} for the symmetric parts are of the exact same form as in the case of a zero magnetic field. Therefore, the analysis of \cite{AG07,BG18} can be immediately applied to~\eref{gen_rel_sym_parts} without any difficulty. On the other hand, care should be taken when generalizing the conclusions of \cite{AG07,BG18} to the relations~\eref{gen_rel_antisym_parts} for the antisymmetric parts. Indeed, it must be noted that the latter differ from the symmetric case~\eref{gen_rel_sym_parts} by a factor $-1$. Therefore, we first treat in subsection~\ref{antisym_subsec} the non-trivial case of the antisymmetric relations~\eref{gen_rel_antisym_parts}. We then apply in subsection~\ref{sym_subsec} the known results of \cite{AG07,BG18} to the symmetric relations~\eref{gen_rel_sym_parts}.


\subsection{The antisymmetric relations \texorpdfstring{\eref{gen_rel_antisym_parts}}{}}\label{antisym_subsec}

Here we investigate the mathematical structure of the set~\eref{gen_rel_antisym_parts} of relations satisfied by the antisymmetric parts $Q^{\mathrm{A}}$.

We first note that~\eref{gen_rel_antisym_parts} generates two distinct series of relations depending on the parity of the index $m$. We have
\begin{eqnarray}
m~\mbox{odd}:\qquad  &&0 = \sum_{j=1}^{n} Q_{\bm{\alpha}_{m} \, , \, \bm{\beta}_{n}}^{(j) \, \mathrm{A}} (\bm{B}) \label{rel_m_odd_antisym}\\
m~\mbox{even}:\qquad  &&Q_{\bm{\alpha}_{m} \, , \, \bm{\beta}_{n}}^{\mathrm{A}} (\bm{B}) = -\frac{1}{2} \sum_{j=1}^{n} Q_{\bm{\alpha}_{m} \, , \, \bm{\beta}_{n}}^{(j) \, \mathrm{A}} (\bm{B}) \label{rel_m_even_antisym}
\end{eqnarray}
where the only difference with the corresponding relations for the symmetric parts obtained from~\eref{gen_rel_sym_parts} lies in the parity of the integer $m$. Despite this difference, we can use the results obtained in \cite{BG18} for a zero magnetic field to show that

\begin{theorem}
The relations~\eref{rel_m_odd_antisym} for any odd index $m$ can be deduced from the relations~\eref{rel_m_even_antisym} corresponding to even indices $m$.
\label{theorem1_antisym}
\end{theorem}

Our strategy to prove this theorem is very similar to the approach used in \cite{BG18}, with differences rising only from the different parity of the index $m$ in~\eref{rel_m_odd_antisym}-\eref{rel_m_even_antisym}. We thus restrict our attention on the specific points of the derivation that depend on the parity of $m$, without going into the details of the steps that are common with the proof exposed in \cite{BG18}.

\begin{proof}
We consider all the relations~\eref{rel_m_odd_antisym}-\eref{rel_m_even_antisym} for a given arbitrary total number $\mathcal{N} = m+n \geqslant 1$ of subscripts and fix an arbitrary odd index $m$ of the form
\begin{eqnarray}
m=m_K \equiv 2K + 1 \qquad\mbox{and}\qquad n=n_K = \mathcal{N} - m_K = \mathcal{N} - 2K - 1 
\label{m+n_K_def_antisym}
\end{eqnarray}
where $K$ is an arbitrary integer such that $0 \leqslant K \leqslant \mathbb{E} \left[ (\mathcal{N}-1) / 2 \right]$. Here and in the sequel $\mathbb{E} (x)$ denotes the integer part of the positive real number $x$, i.e. the natural number $k \geqslant 0$ such that $k \leqslant x < k+1$. We write the relation~\eref{rel_m_odd_antisym} for $m=m_K$ and $n=n_K$, and get
\begin{eqnarray}
0 = \sum_{j=1}^{n_K} Q_{\bm{\alpha}_{m_K} \, , \, \bm{\beta}_{n_K}}^{(j) \, \mathrm{A}} (\bm{B})
\label{rel_m_K_antisym}
\end{eqnarray}
where we recall that the quantities $Q^{(j)}$ are defined by~\eref{Q_j_expr}, i.e. here for the antisymmetric part $Q^{(j) \, \mathrm{A}}$
\begin{eqnarray}
Q_{\bm{\alpha}_{m_K} \, , \, \bm{\beta}_{n_K}}^{(j) \, \mathrm{A}} (\bm{B}) = \sum_{k_1=1}^{n_K} \sum_{k_{2}=1 \atop k_{2} > k_{1}}^{n_K} \cdots \sum_{k_{j}=1 \atop k_{j} > k_{j-1}}^{n_K} Q_{\bm{\alpha}_{m_K} \beta_{k_1} \cdots \beta_{k_{j}} \, , \, (\boldsymbol{\cdot})}^{\mathrm{A}} (\bm{B})
\label{Q_j_antisym_expr}
\end{eqnarray}
for any $1 \leqslant j \leqslant n_K$.

Then we consider all the possible even indices $m=m'_k$ that satisfy $m'_k > m_K = 2K+1$, i.e.
\begin{eqnarray}
m = m'_k\equiv 2 k \qquad\mbox{and}\qquad n = n'_k =\mathcal{N} - m'_k = \mathcal{N} - 2k
\label{prime_indices}
\end{eqnarray}
where $k$ is any integer such that $K+1 \leqslant k \leqslant \mathbb{E} \left( \mathcal{N} / 2 \right)$. Since $m'_k$ is even by construction, then all coefficients $Q_{\bm{\alpha}_{m'_k} \, , \, \bm{\beta}_{n'_k}}^{\mathrm{A}}$ for all possible values of $k$ can be expressed with~\eref{rel_m_even_antisym}.

Finally, we define the partial sums $S_k$ by
\begin{eqnarray}
S_k \equiv \sum_{j=1}^{k} Q_{\bm{\alpha}_{m_K} \, , \, \bm{\beta}_{n_K}}^{(j) \, \mathrm{A}} (\bm{B})
\label{S_k_def}
\end{eqnarray}
where $k$ is an arbitrary integer such that $1 \leqslant k \leqslant n_K$. It is clear that setting $k=n_K$ into~\eref{S_k_def} yields the right-hand side of~\eref{rel_m_K_antisym}. The idea is thus to show that the identity~\eref{rel_m_K_antisym}, i.e. $S_{n_K}=0$, can be deduced from the set of all the relations~\eref{rel_m_even_antisym} for all the possible values of $m>m_K$, which is precisely the content of theorem~\ref{theorem1_antisym}.

The first step is to construct the partial sum~\eref{S_k_def} for an arbitrary even index. We can show by induction that

\begin{proposition*}
For an arbitrary integer $l$ such that $1 \leqslant l \leqslant \mathbb{E} \left[ \left( \mathcal{N}-1 \right) / 2 \right] - K$, the partial sums~\eref{S_k_def} of even indices $k=2l$ can be expressed as
\begin{eqnarray}
S_{2l} = \sum_{j=2l+1}^{n_K} \gamma_{j}^{(2l)} Q_{\bm{\alpha}_{m_K} \, , \, \bm{\beta}_{n_K}}^{(j) \, \mathrm{A}} (\bm{B})
\label{S_2l_expr}
\end{eqnarray}
where the quantities $\gamma_{j}^{(2l)}$ satisfy the recurrence
\begin{eqnarray}
\gamma_{j}^{(2l)} = - \frac{1}{2} \left( \gamma_{2l-1}^{(2l-2)} + 1 \right) \binom{j}{2l-1} + \gamma_{j}^{(2l-2)}
\label{gamma_j_2l_def}
\end{eqnarray}
with $\gamma_j^{(0)}=0$ and the binomial coefficients
\begin{eqnarray}
\binom{p}{q} \equiv \frac{p!}{q! \, (p-q)!} \, .
\label{binom_coef_def}
\end{eqnarray}
\end{proposition*}

A detailed proof of the above proposition can be found in \cite{BG18}. This proof can be readily applied to the antisymmetric quantities $Q_{\bm{\alpha}_{m_K} \, , \, \bm{\beta}_{n_K}}^{(j) \, \mathrm{A}} (\bm{B})$ considered here. Accordingly, we use the expression~\eref{S_2l_expr} of the partial sum $S_{2l}$ for any $1 \leqslant l \leqslant \mathbb{E} \left[ \left( \mathcal{N}-1 \right) / 2 \right] - K$ to prove that $S_{n_K}=0$. We do this by distinguishing the two cases of an odd and of an even total number $\mathcal{N}$ of subscripts.

We first consider an odd $\mathcal{N}$, say $\mathcal{N}=2N+1$ with some integer $N \geqslant 0$. We have in this case $\mathbb{E} \left[ \left( \mathcal{N}-1 \right) / 2 \right] - K = N-K$ and, from the definition~\eref{m+n_K_def_antisym} of $n_K$, $n_K = 2(N-K)$. Therefore, setting $l=N-K$ into~\eref{S_2l_expr} gives $S_{2N-2K}$, i.e. $S_{n_K}$, and we have
\begin{eqnarray*}
S_{n_K} = \sum_{j=n_K+1}^{n_K} \gamma_{j}^{(n_K)} Q_{\bm{\alpha}_{m_K} \, , \, \bm{\beta}_{n_K}}^{(j) \, \mathrm{A}} (\bm{B})
\end{eqnarray*}
that is clearly $S_{n_K} = 0$.

We now consider an even $\mathcal{N}$ of the form $\mathcal{N}=2N$, $N \geqslant 1$, so that $\mathbb{E} \left[ \left( \mathcal{N}-1 \right) / 2 \right] - K = N-K-1$ and $n_K = 2N-2K-1$. Setting now $l=N-K-1$ into~\eref{S_2l_expr} yields
\begin{eqnarray*}
S_{2N-2K-2} = S_{n_K-1} = \gamma_{n_K}^{(n_K-1)} Q_{\bm{\alpha}_{m_K} \, , \, \bm{\beta}_{n_K}}^{(n_K) \, \mathrm{A}} (\bm{B})
\end{eqnarray*}
and thus, in view of the definition~\eref{S_k_def} of the partial sums $S_k$,
\begin{eqnarray}
S_{n_K} = S_{n_K-1} + Q_{\bm{\alpha}_{m_K} \, , \, \bm{\beta}_{n_K}}^{(n_K) \, \mathrm{A}} (\bm{B}) = \left( \gamma_{n_K}^{(n_K-1)} + 1 \right) Q_{\bm{\alpha}_{m_K} \, , \, \bm{\beta}_{n_K}}^{(n_K) \, \mathrm{A}} (\bm{B}) \, .
\label{S_n_K_N_even}
\end{eqnarray}
Using the expression~\eref{Q_j_antisym_expr} of $Q^{(j) \, \mathrm{A}}$ for $j=n_K$ we get
\begin{eqnarray}
Q_{\bm{\alpha}_{m_K} \, , \, \bm{\beta}_{n_K}}^{(n_K) \, \mathrm{A}} (\bm{B}) = \sum_{k_1=1}^{n_K} \sum_{k_{2}=1 \atop k_{2} > k_{1}}^{n_K} \cdots \sum_{k_{n_K}=1 \atop k_{n_K} > k_{n_K-1}}^{n_K} Q_{\bm{\alpha}_{m_K} \beta_{k_1} \cdots \beta_{k_{n_K}}}^{\mathrm{A}} \left( \bm{0} ; \bm{B} \right)
\label{Q_n_K_antisym_expr}
\end{eqnarray}
where we emphasize that the elements of the sum in the right-hand side are statistical cumulants (that indeed do not possess any subscript on the right of the comma). Now, we recall [see~\eref{alpha_beta_def} and~\eref{m+n_K_def_antisym}] that $\bm{\alpha}_{m_K} \equiv \alpha_1 \ldots \alpha_{m_K}$, $\bm{\beta}_{n_K} \equiv \beta_1 \ldots \beta_{n_K}$, $m_K+n_K \equiv \mathcal{N}$ and that we consider an even total number of subscripts, $\mathcal{N}=2N$. We can thus use the relation~\eref{rel_m_even_antisym} for $m=\mathcal{N}$ and $n=\mathcal{N}-m=0$ to rewrite the cumulant $Q_{\bm{\alpha}_{m_K} \beta_{k_1} \cdots \beta_{k_{n_K}}}^{\mathrm{A}} \left( \bm{0} ; \bm{B} \right)$, which is then seen to vanish identically,
\begin{eqnarray}
\fl Q_{\bm{\alpha}_{m_K} \beta_{k_1} \cdots \beta_{k_{n_K}},}^{\mathrm{A}} (\bm{B}) = Q_{\bm{\alpha}_{m_K} \beta_{k_1} \cdots \beta_{k_{n_K}}}^{\mathrm{A}} \left( \bm{0} ; \bm{B} \right) = 0 \qquad\mbox{for even values of}\ \ \mathcal{N} = m_K+n_K\, . \nonumber\\
\label{last_cumulants_antisym_vanish}
\end{eqnarray}
Combining~\eref{S_n_K_N_even}-\eref{Q_n_K_antisym_expr} with~\eref{last_cumulants_antisym_vanish} readily yields, here again, $S_{n_K} = 0$.

Hence we first saw that the total sum $S_{n_K}$ vanishes for $\mathcal{N}$ odd, and that it also vanishes for $\mathcal{N}$ even. We can thus conclude that
\begin{eqnarray}
S_{n_K} = 0 
\label{S_n_K_zero_antisym}
\end{eqnarray}
for any integer $\mathcal{N} \geqslant 1$, which is the desired result.
\end{proof}

Theorem~\ref{theorem1_antisym} shows that, for a fixed $\mathcal{N}=m+n$, the relation obtained from~\eref{gen_rel_antisym_parts} for an arbitrary odd index $m=m_K \equiv 2K+1$ can be deduced from the set of all relations~\eref{gen_rel_antisym_parts} for all even indices $m>m_K$. We can now use this result to prove that

\begin{corollary}
Among all the relations~\eref{gen_rel_antisym_parts}, the set of independent ones is given by~\eref{rel_m_even_antisym} corresponding to even indices $m$.
\label{corollary_antisym}
\end{corollary}

This generalizes the corollary of \cite{BG18} to the antisymmetric relations~\eref{gen_rel_antisym_parts}. Our proof here proceeds along very similar lines.

\begin{proof}
Theorem~\ref{theorem1_antisym} immediately provides the relations~\eref{gen_rel_antisym_parts} that are independent, namely the relations~\eref{rel_m_even_antisym}. This can be seen by considering again a given arbitrary total number $\mathcal{N} = m+n$ of subscripts. We also consider an arbitrary even integer $M$ such that $1 \leqslant M \leqslant \mathcal{N}$. We set $m=M$ into~\eref{rel_m_even_antisym}, and obtain an expression of the coefficient $Q_{\bm{\alpha}_M \, , \, \bm{\beta}_{\mathcal{N}-M}}^{\mathrm{A}}$ in terms of quantities that possess $M+1, \ldots , \mathcal{N}$ subscripts on the left of the comma. We then note that this coefficient does not appear in any of the relations~\eref{rel_m_even_antisym} obtained for any $m>M$. Consequently, the relation~\eref{rel_m_even_antisym} for $m=M$ cannot be deduced from the set of relations~\eref{rel_m_even_antisym} for all $m>M$, which is true for any even integer $M \leqslant \mathcal{N}$. Therefore, the relations~\eref{rel_m_even_antisym} are all independent.
\end{proof}

A direct consequence of the above corollary is thus the following: the antisymmetric part of any even cumulant (and any of their responses) is unambiguously specified by the FR~\eref{fluct_rel}. Therefore, the latter only leaves the antisymmetric part of the odd cumulants (and their responses) undetermined. Furthermore, the independent relations~\eref{rel_m_even_antisym} allow to write the antisymmetric part of any even cumulant (and any of their responses) as a mere linear combination of the antisymmetric part of odd cumulants (and their responses). The coefficients of this linear combination turn out to be coefficients of Euler polynomials. We first recall the standard results regarding these particular polynomials that are required for our subsequent analysis.

The Euler polynomial $E_k(x)$, with $k \geqslant 0$ an integer and $x$ a real number, can be written in terms of its coefficients $e_{i}^{(k)}$ through
\begin{eqnarray}
E_k (x) = \sum_{i=0}^{k} e_{i}^{(k)} x^{i}
\label{Euler_coef_def}
\end{eqnarray}
where it can be shown that $e_k^{(k)}=1$ \cite{GradRyz, AbrSteg}. The constant term $e_{0}^{(k)}$ of $E_k(x)$ can be expressed in terms of particular values of the Euler polynomial through
\begin{eqnarray}
e_{0}^{(k)} = E_k (0) = -E_k(1)
\label{constant_term_Euler}
\end{eqnarray}
for any integer $k \geqslant 0$, as can be seen upon setting $x=0$ into~\eref{Euler_coef_def} and using a well-known property of Euler polynomials \cite{AbrSteg}. Moreover, the constant term is known to satisfy
\begin{eqnarray}
e_0^{(0)}=E_0(0)=1 \qquad \mbox{and}\qquad e_{0}^{(2j)} = E_{2j}(0)=0
\label{e_0_even_vanish}
\end{eqnarray}
for any integer $j \geqslant 1$ \cite{GradRyz, AbrSteg}.  Furthermore, as is for instance shown in~\cite{BG18}, we have the identity
\begin{eqnarray}
e_{k-p}^{(k)} = e_{0}^{(p)} \binom{k}{p}
\label{e_from_e_0}
\end{eqnarray}
for any integers $k$ and $p$ such that $k \geqslant 0$ and $0 \leqslant p \leqslant k$. Equations~\eref{e_0_even_vanish} and~\eref{e_from_e_0} imply that
\begin{eqnarray}
e_{k-2q}^{(k)} = 0
\label{e_k_min_2p_vanish}
\end{eqnarray}
for any integers $k$ and $q$ such that $k \geqslant 2$ and $1 \leqslant q \leqslant \mathbb{E} (k/2)$.

We now use the above well-known properties of Euler polynomials to express the antisymmetric part of the response of any even cumulant only in terms of the antisymmetric part of the response of odd cumulants. We do this by considering again all the relations~\eref{rel_m_even_antisym} for an arbitrarily fixed total number $\mathcal{N}=m+n$ of subscripts, and we show that

\begin{theorem}
Let $\mathcal{N} \geqslant 1$ be an arbitrary total number of subscripts, and $l$ be any integer such that $1 \leqslant l \leqslant \mathbb{E} \left[ (\mathcal{N}+1) / 2 \right]$. Then the antisymmetric part of the response of any even cumulant can be written as the linear combination
\begin{eqnarray}
\fl Q_{\bm{\alpha}_{2\mathcal{N}-2l-2\mathbb{E} \left( \mathcal{N} / 2 \right)} \, , \, \bm{\beta}_{2l+2\mathbb{E} \left( \mathcal{N} / 2 \right)-\mathcal{N}}}^{\mathrm{A}} (\bm{B}) = \sum_{j=1}^{l} e_{0}^{(2j-1)} Q_{\bm{\alpha}_{2\mathcal{N}-2l-2\mathbb{E} \left( \mathcal{N} / 2 \right)} \, , \, \bm{\beta}_{2l+2\mathbb{E} \left( \mathcal{N} / 2 \right)-\mathcal{N}}}^{(2j-1) \, \mathrm{A}} (\bm{B})
\label{even_cumul_antisym_expr}
\end{eqnarray}
where $e_{0}^{(k)}$ denotes the constant term of the Euler polynomial $E_k (x)$.
\label{theorem2_antisym}
\end{theorem}

This generalizes the results obtained in \cite{AG07} in the absence of a magnetic field to the set of relations~\eref{gen_rel_antisym_parts} satisfied by the antisymmetric parts $Q^{\mathrm{A}}$. It is worth emphasizing that the identity~\eref{even_cumul_antisym_expr} involves constant terms $e_{0}^{(k)}$ of Euler polynomials $E_k(x)$, whereas the corresponding expressions in \cite{AG07} make no reference to the latter quantities. Therefore, our proof below of~\eref{even_cumul_antisym_expr} differs significantly from the corresponding derivation in \cite{AG07}.

For clarity, we here rewrite the identity~\eref{even_cumul_antisym_expr} by explicitly distinguishing the two cases of an even and an odd total number $\mathcal{N}$ of subscripts. If $\mathcal{N} \geqslant 2$ is even, we have from~\eref{even_cumul_antisym_expr}
\begin{eqnarray}
\mathcal{N}~\mbox{even}:\qquad  Q_{\bm{\alpha}_{\mathcal{N}-2l} \, , \, \bm{\beta}_{2l}}^{\mathrm{A}} (\bm{B}) = \sum_{j=1}^{l} e_{0}^{(2j-1)} Q_{\bm{\alpha}_{\mathcal{N}-2l} \, , \, \bm{\beta}_{2l}}^{(2j-1) \, \mathrm{A}} (\bm{B})
\label{even_cumul_N_even_antisym_expr}
\end{eqnarray}
for any $1 \leqslant l \leqslant \mathcal{N}/2$, while for an odd $\mathcal{N} \geqslant 1$, the result~\eref{even_cumul_antisym_expr} reads
\begin{eqnarray}
\mathcal{N}~\mbox{odd}:\qquad  Q_{\bm{\alpha}_{\mathcal{N}-2l+1} \, , \, \bm{\beta}_{2l-1}}^{\mathrm{A}} (\bm{B}) = \sum_{j=1}^{l} e_{0}^{(2j-1)} Q_{\bm{\alpha}_{\mathcal{N}-2l+1} \, , \, \bm{\beta}_{2l-1}}^{(2j-1) \, \mathrm{A}} (\bm{B})
\label{even_cumul_N_odd_antisym_expr}
\end{eqnarray}
for any $1 \leqslant l \leqslant (\mathcal{N}+1)/2$.

\begin{proof}
We proceed by induction over the index $l$, and show that the induction hypothesis defined by~\eref{even_cumul_antisym_expr} is true for any integer $l$ such that $1 \leqslant l \leqslant \mathbb{E} \left[ (\mathcal{N}+1) / 2 \right]$. Because of the integer part in~\eref{even_cumul_antisym_expr}, we must distinguish the two cases of an even and of an odd total number $\mathcal{N}$ of subscripts. For concreteness, we treat only the case of $\mathcal{N}$ even. The proof for an odd $\mathcal{N}$ then proceeds along the exact same lines.

Therefore, we consider the case of an even $\mathcal{N}$, say of the form
\begin{eqnarray}
\mathcal{N} = 2N
\label{even_tot_nb_sub}
\end{eqnarray}
where $N \geqslant 1$ is an arbitrary integer. We consider all possible relations~\eref{rel_m_even_antisym} for this fixed even $\mathcal{N}=m+n$. We hence first set $m=\mathcal{N}$, i.e. $n=\mathcal{N}-m=0$, into~\eref{rel_m_even_antisym} to get
\begin{eqnarray}
Q_{\bm{\alpha}_{\mathcal{N}}}^{\mathrm{A}} \left( \bm{0} ; \bm{B} \right) = 0 \qquad\mbox{for even values of}\ \ \mathcal{N} \, .
\label{even_cumul_antisym_zero}
\end{eqnarray}
This shows that the antisymmetric part of any even cumulant vanishes, which could already be seen on the identity~\eref{parity_m_cumu}. We then consider the relations~\eref{rel_m_even_antisym} corresponding to $m=\mathcal{N}-2, \mathcal{N}-4, \ldots$ to show by induction that the hypothesis~\eref{even_cumul_antisym_expr}, i.e. equivalently the hypothesis~\eref{even_cumul_N_even_antisym_expr} since we have an even $\mathcal{N}$ here, is true for any integer $l$ such that $1 \leqslant l \leqslant \mathcal{N}/2$.

We first initialize the recurrence by constructing the quantity $Q_{\bm{\alpha}_{\mathcal{N}-2} \, , \, \bm{\beta}_{2}}^{\mathrm{A}} (\bm{B})$. We hence set $m=\mathcal{N}-2$ and $n=\mathcal{N}-m=2$ into~\eref{rel_m_even_antisym} to get
\begin{eqnarray}
Q_{\bm{\alpha}_{\mathcal{N}-2} \, , \, \bm{\beta}_{2}}^{\mathrm{A}} (\bm{B}) = - \frac{1}{2} \left[ Q_{\bm{\alpha}_{\mathcal{N}-2} \, , \, \bm{\beta}_{2}}^{(1) \, \mathrm{A}} (\bm{B}) + Q_{\bm{\alpha}_{\mathcal{N}-2} \, , \, \bm{\beta}_{2}}^{(2) \, \mathrm{A}} (\bm{B}) \right] \, .
\label{Q_N-2_2_antisym_expr}
\end{eqnarray}
We recall that the quantities $Q^{(j)}$ are defined by~\eref{Q_j_expr}, so that we have here for the antisymmetric part $Q^{(2) \, \mathrm{A}}$
\begin{eqnarray*}
Q_{\bm{\alpha}_{\mathcal{N}-2} \, , \, \bm{\beta}_{2}}^{(2) \, \mathrm{A}} (\bm{B}) = \sum_{k_1=1}^{2} \sum_{k_{2}=1 \atop k_{2} > k_{1}}^{2} Q_{\bm{\alpha}_{\mathcal{N}-2} \beta_{k_1} \beta_{k_2}}^{\mathrm{A}} \left( \bm{0} ; \bm{B} \right)
\end{eqnarray*}
where we emphasize that the terms of the sum in the right-hand side possess $\mathcal{N}$ subscripts on the left of the comma and indeed correspond to $\mathcal{N}$th cumulants (that do not have any subscripts on the right of the comma). We can thus rewrite them in view of~\eref{even_cumul_antisym_zero} and readily get
\begin{eqnarray}
Q_{\bm{\alpha}_{\mathcal{N}-2} \, , \, \bm{\beta}_{2}}^{(2) \, \mathrm{A}} (\bm{B}) = 0 \qquad\mbox{for even values of}\ \ \mathcal{N} \, .
\label{Q_N-2_2_2_antisym_expr}
\end{eqnarray}
Combining~\eref{Q_N-2_2_antisym_expr} with~\eref{Q_N-2_2_2_antisym_expr}, and recalling that $e_0^{(1)}=-1/2$ yields at once
\begin{eqnarray}
Q_{\bm{\alpha}_{\mathcal{N}-2} \, , \, \bm{\beta}_{2}}^{\mathrm{A}} (\bm{B}) = e_0^{(1)} Q_{\bm{\alpha}_{\mathcal{N}-2} \, , \, \bm{\beta}_{2}}^{(1) \, \mathrm{A}} (\bm{B}) \, ,
\label{Q_N-2_2_antisym_expr_ind}
\end{eqnarray}
which hence shows that the induction hypothesis~\eref{even_cumul_N_even_antisym_expr} is indeed true for $l=1$.

We now assume that the induction hypothesis~\eref{even_cumul_N_even_antisym_expr} is true for some integer $l-1$ such that $1 \leqslant l-1 < \mathcal{N}/2$, and thus must show that~\eref{even_cumul_N_even_antisym_expr} remains true for the integer $l$. We first set $m=\mathcal{N}-2l$ and $n=\mathcal{N}-m=2l$ into~\eref{rel_m_even_antisym} to get the quantity $Q_{\bm{\alpha}_{\mathcal{N}-2l} \, , \, \bm{\beta}_{2l}}^{\mathrm{A}} (\bm{B})$, which we write in the form
\begin{eqnarray}
Q_{\bm{\alpha}_{\mathcal{N}-2l} \, , \, \bm{\beta}_{2l}}^{\mathrm{A}} (\bm{B}) = -\frac{1}{2} \left[ \sum_{j=1}^{l} Q_{\bm{\alpha}_{\mathcal{N}-2l} \, , \, \bm{\beta}_{2l}}^{(2j-1) \, \mathrm{A}} (\bm{B}) + \sum_{j=1}^{l} Q_{\bm{\alpha}_{\mathcal{N}-2l} \, , \, \bm{\beta}_{2l}}^{(2j) \, \mathrm{A}} (\bm{B}) \right] \, .
\label{Q_N-2l_2l_antisym_def}
\end{eqnarray}
We first note that in view of~\eref{Q_j_expr} we have here for the antisymmetric part $Q^{(2l) \, \mathrm{A}}$
\begin{eqnarray*}
Q_{\bm{\alpha}_{\mathcal{N}-2l} \, , \, \bm{\beta}_{2l}}^{(2l) \, \mathrm{A}} (\bm{B}) = \sum_{k_1=1}^{2l} \sum_{k_{2}=1 \atop k_{2} > k_{1}}^{2l} \cdots \sum_{k_{2l}=1 \atop k_{2l} > k_{2l-1}}^{2l} Q_{\bm{\alpha}_{\mathcal{N}-2l} \beta_{k_1} \cdots \beta_{k_{2l}}}^{\mathrm{A}} \left( \bm{0} ; \bm{B} \right)
\end{eqnarray*}
where the terms of the sum in the right-hand side possess $\mathcal{N}$ subscripts on the left of the comma and indeed correspond to $\mathcal{N}$th cumulants, which vanish according to~\eref{even_cumul_antisym_zero}. We hence have
\begin{eqnarray}
Q_{\bm{\alpha}_{\mathcal{N}-2l} \, , \, \bm{\beta}_{2l}}^{(2l) \, \mathrm{A}} (\bm{B}) = 0 \qquad\mbox{for even values of}\ \ \mathcal{N}
\label{Q_N-2l_2l_2l_antisym_expr}
\end{eqnarray}
so that~\eref{Q_N-2l_2l_antisym_def} reads
\begin{eqnarray}
Q_{\bm{\alpha}_{\mathcal{N}-2l} \, , \, \bm{\beta}_{2l}}^{\mathrm{A}} (\bm{B}) = -\frac{1}{2} \left[ \sum_{j=1}^{l} Q_{\bm{\alpha}_{\mathcal{N}-2l} \, , \, \bm{\beta}_{2l}}^{(2j-1) \, \mathrm{A}} (\bm{B}) + \sum_{j=1}^{l-1} Q_{\bm{\alpha}_{\mathcal{N}-2l} \, , \, \bm{\beta}_{2l}}^{(2j) \, \mathrm{A}} (\bm{B}) \right]
\label{Q_N-2l_2l_antisym_expr}
\end{eqnarray}
where the second sum in the right-hand side here stops at $l-1$. We now use our induction hypothesis~\eref{even_cumul_N_even_antisym_expr}, which we assume is true and expresses the quantity $Q_{\bm{\alpha}_{\mathcal{N}-2i} \, , \, \bm{\beta}_{2i}}^{\mathrm{A}} (\bm{B})$ for any integer $i$ such that $1 \leqslant i \leqslant l-1$, to rewrite the second sum in the right-hand side of~\eref{Q_N-2l_2l_antisym_expr}.

To do this, we must construct explicitly the quantity $Q_{\bm{\alpha}_{\mathcal{N}-2l} \, , \, \bm{\beta}_{2l}}^{(2j) \, \mathrm{A}} (\bm{B})$, for any $1 \leqslant j \leqslant l-1$, by means of our induction hypothesis~\eref{even_cumul_N_even_antisym_expr}. First, we have in view of~\eref{Q_j_expr}
\begin{eqnarray}
Q_{\bm{\alpha}_{\mathcal{N}-2l} \, , \, \bm{\beta}_{2l}}^{(2j) \, \mathrm{A}} (\bm{B}) = \sum_{k_1=1}^{2l} \sum_{k_{2}=1 \atop k_{2} > k_{1}}^{2l} \cdots \sum_{k_{2j}=1 \atop k_{2j} > k_{2j-1}}^{2l} Q_{\bm{\alpha}_{\mathcal{N}-2l} \beta_{k_1} \cdots \beta_{k_{2j}} \, , \, (\boldsymbol{\cdot})}^{\mathrm{A}} (\bm{B})
\label{Q_N-2l_2l_2j_antisym_expr}
\end{eqnarray}
where we note that the terms of the sum in the right-hand side possess $\mathcal{N}-2(l-j)$ subscripts on the left of the comma. Since it is clear that the integer $j$ is such that $1 \leqslant l-j \leqslant l-1$, then we can use our induction hypothesis~\eref{even_cumul_N_even_antisym_expr} (upon the substitution $l \to l-j$) to rewrite the terms of the sum in the right-hand side of~\eref{Q_N-2l_2l_2j_antisym_expr} and we have
\begin{eqnarray}
Q_{\bm{\alpha}_{\mathcal{N}-2l} \beta_{k_1} \cdots \beta_{k_{2j}} \, , \, (\boldsymbol{\cdot})}^{\mathrm{A}} (\bm{B}) = \sum_{p=1}^{l-j} e_{0}^{(2p-1)} Q_{\bm{\alpha}_{\mathcal{N}-2l} \beta_{k_1} \cdots \beta_{k_{2j}} \, , \, (\boldsymbol{\cdot})}^{(2p-1) \, \mathrm{A}} (\bm{B})
\label{Q_N-2l_2j_antisym_expr}
\end{eqnarray}
for any $1 \leqslant j \leqslant l-1$. We then express $Q^{(2p-1) \, \mathrm{A}}$ by means of the definition~\eref{Q_j_expr}, i.e.
\begin{eqnarray}
\fl Q_{\bm{\alpha}_{\mathcal{N}-2l} \beta_{k_1} \cdots \beta_{k_{2j}} \, , \, (\boldsymbol{\cdot})}^{(2p-1) \, \mathrm{A}} (\bm{B}) = \sum_{k'_1=1 \atop k'_{1} \neq k_{i}}^{2l} \sum_{k'_{2}=1 \atop {k'_{2} > k'_{1} \atop k'_{2} \neq k_{i}}}^{2l} \cdots \sum_{k'_{2p-1}=1 \atop {k'_{2p-1} > k'_{2p-2} \atop k'_{2p-1} \neq k_{i}}}^{2l} Q_{\bm{\alpha}_{\mathcal{N}-2l} \beta_{k_1} \cdots \beta_{k_{2j}} \beta_{k'_1} \cdots \beta_{k'_{2p-1}} \, , \, (\boldsymbol{\cdot})}^{\mathrm{A}} (\bm{B})
\label{Q_N-2l_2p-1_antisym_expr}
\end{eqnarray}
where $k' \neq k_i$ merely means $k' \neq k_1, \ldots , k_{2j}$. We emphasize that the sums over $k'_1, \ldots , k'_{2p-1}$ in the right-hand side of~\eref{Q_N-2l_2p-1_antisym_expr} must indeed run from 1 to $2l$, and not $2l-2j$ as the definition~\eref{Q_j_expr} might suggest at first sight. It is here important to remember from~\eref{Q_N-2l_2l_2j_antisym_expr}-\eref{Q_N-2l_2j_antisym_expr} that the quantity $Q_{\bm{\alpha}_{\mathcal{N}-2l} \beta_{k_1} \cdots \beta_{k_{2j}} \, , \, (\boldsymbol{\cdot})}^{(2p-1) \, \mathrm{A}} (\bm{B})$ contains, on the right of the comma, subscripts $\beta_k$ where $k$ can take any value between 1 and $2l$, under the condition that $k \neq k_1, \ldots , k_{2j}$. Therefore, we must indeed have in~\eref{Q_N-2l_2p-1_antisym_expr} sums over $k'$ from 1 to $2l$, under the additional conditions that $k' \neq k_1 , \ldots , k_{2j}$. We now substitute~\eref{Q_N-2l_2j_antisym_expr}-\eref{Q_N-2l_2p-1_antisym_expr} into~\eref{Q_N-2l_2l_2j_antisym_expr}, and write the resulting expression of $Q_{\bm{\alpha}_{\mathcal{N}-2l} \, , \, \bm{\beta}_{2l}}^{(2j) \, \mathrm{A}} (\bm{B})$ in the form
\begin{eqnarray}
Q_{\bm{\alpha}_{\mathcal{N}-2l} \, , \, \bm{\beta}_{2l}}^{(2j) \, \mathrm{A}} (\bm{B}) = \sum_{p=1}^{l-j} e_{0}^{(2p-1)} \mathcal{S}_{p}^{(j)}
\label{Q_N-2l_2l_2j_antisym_temp_expr}
\end{eqnarray}
where the quantity $\mathcal{S}_{p}^{(j)}$ is defined by
\begin{eqnarray}
\fl \mathcal{S}_{p}^{(j)} \equiv \sum_{k_1=1}^{2l} \sum_{k_{2}=1 \atop k_{2} > k_{1}}^{2l} \cdots \sum_{k_{2j}=1 \atop k_{2j} > k_{2j-1}}^{2l} \sum_{k'_1=1 \atop k'_{1} \neq k_{i}}^{2l} \sum_{k'_{2}=1 \atop {k'_{2} > k'_{1} \atop k'_{2} \neq k_{i}}}^{2l} \cdots \sum_{k'_{2p-1}=1 \atop {k'_{2p-1} > k'_{2p-2} \atop k'_{2p-1} \neq k_{i}}}^{2l} Q_{\bm{\alpha}_{\mathcal{N}-2l} \beta_{k_1} \cdots \beta_{k_{2j}} \beta_{k'_1} \cdots \beta_{k'_{2p-1}} \, , \, (\boldsymbol{\cdot})}^{\mathrm{A}} (\bm{B}) \, . \nonumber\\
\label{S_p_j_def}
\end{eqnarray}

We now identify in $\mathcal{S}_{p}^{(j)}$ the coefficient of the term $Q_{\bm{\alpha}_{\mathcal{N}-2l} \beta_{h_1} \cdots \beta_{h_{2j+2p-1}} \, , \, (\boldsymbol{\cdot})}^{\mathrm{A}} (\bm{B})$ with $h_1 < \ldots < h_{2j+2p-1}$. We first note that the latter condition defines a set $\mathscr{S}_{2j+2p-1} \equiv \{ h_1,\ldots,h_{2j+2p-1} \}$ of $2j+2p-1$ elements. The term $Q_{\bm{\alpha}_{\mathcal{N}-2l} \beta_{h_1} \cdots \beta_{h_{2j+2p-1}} \, , \, (\boldsymbol{\cdot})}^{\mathrm{A}} (\bm{B})$ then rises from all samples (which are by definition ordered collections of elements of a set) $\{ k_1,\ldots,k_{2j},k'_1,\ldots,k'_{2p-1} \}$ of $\mathscr{S}_{2j+2p-1}$ that are required to satisfy the constraints $k'_1 , \ldots , k'_{2p-1} \neq k_1, \ldots , k_{2j}$, $k_1 < \ldots < k_{2j}$ and $k'_1 < \ldots < k'_{2p-1}$. Because of the latter constraints, our task thus merely consists in counting the total number of \textit{subsets} (which are \textit{unordered} collections of elements of a set) $\{ k'_1,\ldots,k'_{2p-1} \}$ of the set $\mathscr{S}_{2j+2p-1}$. It is in particular worth noting that fixing $\{ k'_1,\ldots,k'_{2p-1} \}$ to be a definite subset of $\mathscr{S}_{2j+2p-1}$ selects a unique subset $\{ k_1,\ldots,k_{2j} \}$ of $\mathscr{S}_{2j+2p-1}$. This total number of subsets is known to be $\binom{2j+2p-1}{2p-1}$ (see e.g. theorem~4.1 in~\cite{Rys}). This shows that $\mathcal{S}_{p}^{(j)}$ contains exactly $\binom{2j+2p-1}{2p-1}$ terms $Q_{\bm{\alpha}_{\mathcal{N}-2l} \beta_{h_1} \cdots \beta_{h_{2j+2p-1}} \, , \, (\boldsymbol{\cdot})}^{\mathrm{A}} (\bm{B})$ with $h_1 < \ldots < h_{2j+2p-1}$, so that~\eref{S_p_j_def} reads, after recognizing the quantity $Q_{\bm{\alpha}_{\mathcal{N}-2l} \, , \, \bm{\beta}_{2l}}^{(2j+2p-1) \, \mathrm{A}} (\bm{B})$ as given by~\eref{Q_j_expr},
\begin{eqnarray}
\mathcal{S}_{p}^{(j)} = \binom{2j+2p-1}{2p-1} \, Q_{\bm{\alpha}_{\mathcal{N}-2l} \, , \, \bm{\beta}_{2l}}^{(2j+2p-1) \, \mathrm{A}} (\bm{B}) \, .
\label{S_p_j_expr}
\end{eqnarray}
Substituting this expression of $\mathcal{S}_{p}^{(j)}$ into~\eref{Q_N-2l_2l_2j_antisym_temp_expr} hence yields
\begin{eqnarray}
Q_{\bm{\alpha}_{\mathcal{N}-2l} \, , \, \bm{\beta}_{2l}}^{(2j) \, \mathrm{A}} (\bm{B}) = \sum_{p=1}^{l-j} e_{0}^{(2p-1)} \binom{2j+2p-1}{2p-1} \, Q_{\bm{\alpha}_{\mathcal{N}-2l} \, , \, \bm{\beta}_{2l}}^{(2j+2p-1) \, \mathrm{A}} (\bm{B})
\label{Q_N-2l_2l_2j_antisym_fin_expr}
\end{eqnarray}
for any $1 \leqslant j \leqslant l-1$. We now use this expression of $Q_{\bm{\alpha}_{\mathcal{N}-2l} \, , \, \bm{\beta}_{2l}}^{(2j) \, \mathrm{A}} (\bm{B})$ to compute the second sum in the right-hand side of~\eref{Q_N-2l_2l_antisym_expr}, for which we introduce the notation
\begin{eqnarray}
\widetilde{S}_l \equiv \sum_{j=1}^{l-1} Q_{\bm{\alpha}_{\mathcal{N}-2l} \, , \, \bm{\beta}_{2l}}^{(2j) \, \mathrm{A}} (\bm{B}) \, .
\label{S_tilde_l_def}
\end{eqnarray}

We first substitute~\eref{Q_N-2l_2l_2j_antisym_fin_expr} into~\eref{S_tilde_l_def} to get
\begin{eqnarray}
\widetilde{S}_l = \sum_{j=1}^{l-1} \sum_{p=1}^{l-j} e_{0}^{(2p-1)} \binom{2j+2p-1}{2p-1} \, Q_{\bm{\alpha}_{\mathcal{N}-2l} \, , \, \bm{\beta}_{2l}}^{(2j+2p-1) \, \mathrm{A}} (\bm{B}) \, ,
\label{S_tilde_l_expr}
\end{eqnarray}
which, in view of the first sum in the right-hand side of~\eref{Q_N-2l_2l_antisym_expr}, we want to rewrite as a single sum, with respect to an index $j'$ say, of quantities $Q_{\bm{\alpha}_{\mathcal{N}-2l} \, , \, \bm{\beta}_{2l}}^{(2j'-1) \, \mathrm{A}} (\bm{B})$. By setting $j'=j+p$ into~\eref{S_tilde_l_expr}, we can thus write $\widetilde{S}_l$ as the sum
\begin{eqnarray}
\widetilde{S}_l = \sum_{j'=2}^{l} \Gamma_{j'} \, Q_{\bm{\alpha}_{\mathcal{N}-2l} \, , \, \bm{\beta}_{2l}}^{(2j'-1) \, \mathrm{A}} (\bm{B})
\label{S_tilde_l_desired_expr}
\end{eqnarray}
where the coefficients $\Gamma_{j'}$ are to be determined. It is first worth noting that the lower and upper limits of the sum in~\eref{S_tilde_l_desired_expr} are indeed right, for we see from~\eref{S_tilde_l_expr} that $\widetilde{S}_l$ consists in a linear combination of the quantities $Q^{(3) \, \mathrm{A}} , \ldots , Q^{(2l-1) \, \mathrm{A}}$. Now, we must identify on~\eref{S_tilde_l_expr} the coefficient $\Gamma_{j'}$ of the term $Q_{\bm{\alpha}_{\mathcal{N}-2l} \, , \, \bm{\beta}_{2l}}^{(2j'-1) \, \mathrm{A}} (\bm{B})$, for an arbitrary $2 \leqslant j' \leqslant l$. Because $j'=j+p$, the latter term rises from the following combinations of the indices $j$ and $p$:
\begin{eqnarray*}
\left\{ j \, , \, p \right\} = \left\{ 1 \, , \, j'-1 \right\}, \left\{ 2 \, , \, j'-2 \right\}, \ldots , \left\{ j'-1 \, , \, 1 \right\} \, .
\end{eqnarray*}
Therefore, the coefficient $\Gamma_{j'}$ is given by (with the index $k=p$)
\begin{eqnarray}
\Gamma_{j'} = \sum_{k=1}^{j'-1} e_{0}^{(2k-1)} \binom{2j'-1}{2k-1}
\label{Gamma_e_0_expr}
\end{eqnarray}
that is, combining~\eref{Gamma_e_0_expr} with the property~\eref{e_from_e_0} of coefficients of Euler polynomials,
\begin{eqnarray}
\Gamma_{j'} = \sum_{k=1}^{j'-1} e_{2k}^{(2j'-1)} \, .
\label{Gamma_e_expr}
\end{eqnarray}
We then combine~\eref{S_tilde_l_desired_expr} with~\eref{Gamma_e_expr}, relabel the dummy index $j'$ by $j$, and remember the definition~\eref{S_tilde_l_def} of the sum $\widetilde{S}_l$ to get the identity
\begin{eqnarray}
\widetilde{S}_l  \equiv \sum_{j=1}^{l-1} Q_{\bm{\alpha}_{\mathcal{N}-2l} \, , \, \bm{\beta}_{2l}}^{(2j) \, \mathrm{A}} (\bm{B}) = \sum_{j=2}^{l} \left( \sum_{k=1}^{j-1} e_{2k}^{(2j-1)} \right) \, Q_{\bm{\alpha}_{\mathcal{N}-2l} \, , \, \bm{\beta}_{2l}}^{(2j-1) \, \mathrm{A}} (\bm{B}) \, .
\label{second_sum_expr}
\end{eqnarray}

We now substitute the result~\eref{second_sum_expr} into the expression~\eref{Q_N-2l_2l_antisym_expr} of $Q_{\bm{\alpha}_{\mathcal{N}-2l} \, , \, \bm{\beta}_{2l}}^{\mathrm{A}} (\bm{B})$, which yields
\begin{eqnarray}
\fl Q_{\bm{\alpha}_{\mathcal{N}-2l} \, , \, \bm{\beta}_{2l}}^{\mathrm{A}} (\bm{B}) = -\frac{1}{2} \, Q_{\bm{\alpha}_{\mathcal{N}-2l} \, , \, \bm{\beta}_{2l}}^{(1) \, \mathrm{A}} (\bm{B}) -\frac{1}{2} \sum_{j=2}^{l} \left( 1 + \sum_{k=1}^{j-1} e_{2k}^{(2j-1)} \right) Q_{\bm{\alpha}_{\mathcal{N}-2l} \, , \, \bm{\beta}_{2l}}^{(2j-1) \, \mathrm{A}} (\bm{B}) \, .
\label{Q_N-2l_2l_antisym_temp_expr}
\end{eqnarray}
Noting that the coefficients of Euler polynomials are known to satisfy $e_{0}^{(1)}=-1/2$ and $e_{p}^{(p)}=1$ for any $p \geqslant 0$, we can thus rewrite~\eref{Q_N-2l_2l_antisym_temp_expr} as
\begin{eqnarray}
\fl Q_{\bm{\alpha}_{\mathcal{N}-2l} \, , \, \bm{\beta}_{2l}}^{\mathrm{A}} (\bm{B}) = e_{0}^{(1)} \, Q_{\bm{\alpha}_{\mathcal{N}-2l} \, , \, \bm{\beta}_{2l}}^{(1) \, \mathrm{A}} (\bm{B}) \nonumber \\[0.25cm]
-\frac{1}{2} \sum_{j=2}^{l} \left( e_{2j-1}^{(2j-1)} + \sum_{k=1}^{j-1} e_{2k}^{(2j-1)} \right) Q_{\bm{\alpha}_{\mathcal{N}-2l} \, , \, \bm{\beta}_{2l}}^{(2j-1) \, \mathrm{A}} (\bm{B}) \, .
\label{Q_N-2l_2l_antisym_from_e_expr}
\end{eqnarray}
Now, in view of the property~\eref{e_k_min_2p_vanish} we have
\begin{eqnarray*}
e_{2j-1}^{(2j-1)} + \sum_{k=1}^{j-1} e_{2k}^{(2j-1)} = \sum_{i=0}^{2j-1} e_{i}^{(2j-1)} - e_{0}^{(2j-1)}
\end{eqnarray*}
where we recognize in the right-hand side the expression~\eref{Euler_coef_def} of $E_{2j-1} (1)$. We hence get, in view of~\eref{constant_term_Euler},
\begin{eqnarray}
e_{2j-1}^{(2j-1)} + \sum_{k=1}^{j-1} e_{2k}^{(2j-1)} = -E_{2j-1} (0) - e_{0}^{(2j-1)} = - 2 e_{0}^{(2j-1)} \, .
\label{Euler_Q_N-2l_2l}
\end{eqnarray}
Finally, substituting the identity~\eref{Euler_Q_N-2l_2l} into~\eref{Q_N-2l_2l_antisym_from_e_expr} precisely yields the desired expression \eref{even_cumul_N_even_antisym_expr} of $Q_{\bm{\alpha}_{\mathcal{N}-2l} \, , \, \bm{\beta}_{2l}}^{\mathrm{A}} (\bm{B})$. This shows that the induction hypothesis~\eref{even_cumul_N_even_antisym_expr} is indeed true for the integer $l$.

Therefore, we showed (i) that the induction hypothesis~\eref{even_cumul_N_even_antisym_expr} is true for $l=1$, and (ii) that if~\eref{even_cumul_N_even_antisym_expr} is true for some integer $l-1$, then it remains true for the integer $l$. It must thus be true for any integer $1 \leqslant l \leqslant \mathcal{N}/2$. This shows that the result~\eref{even_cumul_antisym_expr} is indeed true for an arbitrary even total number $\mathcal{N}$ of subscripts.

The proof then proceeds in the exact same way in the case of an odd $\mathcal{N}$, which hence shows that theorem~\ref{theorem2_antisym} indeed holds.
\end{proof}

We investigated in this subsection how results obtained in the absence of a magnetic field \cite{AG07,BG18} (and that immediately apply to the relations~\eref{gen_rel_sym_parts} for the symmetric parts, as is discussed in next subsection~\ref{sym_subsec}) extend to the relations~\eref{gen_rel_antisym_parts} for the antisymmetric parts. We first reached, with theorem~\ref{theorem1_antisym} and corollary~\ref{corollary_antisym} above, conclusions that are very similar to \cite{BG18} regarding the independent relations of the set~\eref{gen_rel_antisym_parts}. We then obtained with theorem~\ref{theorem2_antisym} the expression~\eref{even_cumul_antisym_expr} of the antisymmetric part of the response of any even cumulant in terms of the antisymmetric part of the response of odd cumulants only, which generalizes the findings of \cite{AG07} to the antisymmetric relations~\eref{gen_rel_antisym_parts}. We stress that an interesting feature of the identity~\eref{even_cumul_antisym_expr} as compared to the corresponding relation of \cite{AG07} is its explicit connection with the coefficients $e_{0}^{(k)}$ of Euler polynomials.

We now perform the same analysis regarding the relations~\eref{gen_rel_sym_parts} for the symmetric parts.


\subsection{The symmetric relations \texorpdfstring{\eref{gen_rel_sym_parts}}{}}\label{sym_subsec}

Here we consider the relations~\eref{gen_rel_sym_parts} satisfied by the symmetric parts $Q^{\mathrm{S}}$. Our analysis is now greatly simplified upon noting that the relations~\eref{gen_rel_sym_parts} are of the exact same form as in the case of a zero magnetic field.

Here again we distinguish the two series of relations obtained from~\eref{gen_rel_sym_parts} depending on the parity of the index $m$, i.e.
\begin{eqnarray}
m~\mbox{even}:\qquad  &&0 = \sum_{j=1}^{n} Q_{\bm{\alpha}_{m} \, , \, \bm{\beta}_{n}}^{(j) \, \mathrm{S}} (\bm{B}) \label{rel_m_even_sym}\\
m~\mbox{odd}:\qquad  &&Q_{\bm{\alpha}_{m} \, , \, \bm{\beta}_{n}}^{\mathrm{S}} (\bm{B}) = -\frac{1}{2} \sum_{j=1}^{n} Q_{\bm{\alpha}_{m} \, , \, \bm{\beta}_{n}}^{(j) \, \mathrm{S}} (\bm{B}) \, , \label{rel_m_odd_sym}
\end{eqnarray}
which are exactly the same relations that arise in the case of a zero magnetic field. Therefore, the results of \cite{BG18} can be immediately applied to the relations~\eref{rel_m_even_sym}-\eref{rel_m_odd_sym} and we have

\begin{theorem}
The relations~\eref{rel_m_even_sym} for any even index $m$ can be deduced from the relations~\eref{rel_m_odd_sym} corresponding to odd indices $m$.
\label{theorem1_sym}
\end{theorem}

\noindent A direct consequence is thus

\begin{corollary}
Among all the relations~\eref{gen_rel_sym_parts}, the set of independent ones is given by~\eref{rel_m_odd_sym} corresponding to odd indices $m$.
\label{corollary_sym}
\end{corollary}

The detailed proofs of theorem~\ref{theorem1_sym} and corollary~\ref{corollary_sym} can be found in \cite{BG18}, and are thus not reproduced here.

Corollary~\ref{corollary_sym} readily shows that, contrary to the antisymmetric parts, the FR~\eref{fluct_rel} now unambiguously specifies the symmetric part of the response of any \textit{odd} cumulant. The latter can be expressed as a linear combination of the symmetric part of the response of even cumulants only. Similarly to the corresponding result for the antisymmetric parts (i.e. theorem~\ref{theorem2_antisym} above), the coefficients of this linear combination are again coefficients of Euler polynomials and we have

\begin{theorem}
Let $\mathcal{N} \geqslant 1$ be an arbitrary total number of subscripts, and $l$ be any integer such that $1 \leqslant l \leqslant \mathbb{E} \left( \mathcal{N} / 2 \right)$. Then the symmetric part of the response of any odd cumulant can be written as the linear combination
\begin{eqnarray}
\fl Q_{\bm{\alpha}_{2\mathcal{N}-2l-2\mathbb{E} \left[ (\mathcal{N}+1) / 2 \right] + 1} \, , \, \bm{\beta}_{2l+2\mathbb{E} \left[ (\mathcal{N}+1) / 2 \right]-\mathcal{N} - 1}}^{\mathrm{S}} (\bm{B}) \nonumber \\[0.2cm]
= \sum_{j=1}^{l} e_{0}^{(2j-1)} Q_{\bm{\alpha}_{2\mathcal{N}-2l-2\mathbb{E} \left[ (\mathcal{N}+1) / 2 \right] + 1} \, , \, \bm{\beta}_{2l+2\mathbb{E} \left[ (\mathcal{N}+1) / 2 \right]-\mathcal{N} - 1}}^{(2j-1) \, \mathrm{S}} (\bm{B})
\label{odd_cumul_sym_expr}
\end{eqnarray}
where $e_{0}^{(k)}$ denotes the constant term of the Euler polynomial $E_k(x)$.
\label{theorem2_sym}
\end{theorem}

We prove this result with the exact same reasoning as we used regarding theorem~\ref{theorem2_antisym}, which we recall expresses the antisymmetric part of the response of any even cumulant.

It is worth rewriting the identity~\eref{odd_cumul_sym_expr} by explicitly distinguishing the two cases of an odd and an even total number $\mathcal{N}$ of subscripts. First, for an odd $\mathcal{N} \geqslant 1$, the result~\eref{odd_cumul_sym_expr} yields
\begin{eqnarray}
\mathcal{N}~\mbox{odd}:\qquad  Q_{\bm{\alpha}_{\mathcal{N}-2l} \, , \, \bm{\beta}_{2l}}^{\mathrm{S}} (\bm{B}) = \sum_{j=1}^{l} e_{0}^{(2j-1)} Q_{\bm{\alpha}_{\mathcal{N}-2l} \, , \, \bm{\beta}_{2l}}^{(2j-1) \, \mathrm{S}} (\bm{B})
\label{odd_cumul_N_odd_sym_expr}
\end{eqnarray}
for any $1 \leqslant l \leqslant (\mathcal{N}-1)/2$, while for an even $\mathcal{N} \geqslant 2$ we get from~\eref{odd_cumul_sym_expr}
\begin{eqnarray}
\mathcal{N}~\mbox{even}:\qquad  Q_{\bm{\alpha}_{\mathcal{N}-2l+1} \, , \, \bm{\beta}_{2l-1}}^{\mathrm{S}} (\bm{B}) = \sum_{j=1}^{l} e_{0}^{(2j-1)} Q_{\bm{\alpha}_{\mathcal{N}-2l+1} \, , \, \bm{\beta}_{2l-1}}^{(2j-1) \, \mathrm{S}} (\bm{B})
\label{odd_cumul_N_even_sym_expr}
\end{eqnarray}
for any $1 \leqslant l \leqslant \mathcal{N}/2$.

Now, as we already stated above, the relations~\eref{rel_m_even_sym}-\eref{rel_m_odd_sym} are of the exact same form as in the case of a zero magnetic field. Therefore, the expression~\eref{odd_cumul_sym_expr} of the symmetric part of the response of any odd cumulant must be equivalent to the corresponding results obtained in \cite{AG07}. This equivalence allows us to infer a general identity satisfied by the constant terms $e_{0}$ of Euler polynomials, as we now discuss.

\begin{table}[ht]
\centering
\caption{Correspondence between the symmetric quantities $Q^{\mathrm{S}} (\bm{B})$ and $Q^{(j) \, \mathrm{S}} (\bm{B})$ and the related quantities in \cite{AG07} (where we recall that $\bm{B}=\bm{0}$).}
\vskip 0.2 cm
\begin{tabular}{ |M{3cm}|M{3cm}|N  }
 \hline
 Symmetric quantity & Corresponding quantity in \cite{AG07} & \\[0.5cm]
 \hline\hline
 $Q_{\bm{\alpha}_{m} \, , \, \bm{\beta}_{n}}^{\mathrm{S}} (\bm{B})$ & $Q_{\bm{\alpha}_{m} \, , \, \bm{\beta}_{n}}^{(m,m+n)}$ & \\[0.4cm]
 $Q_{\bm{\alpha}_{m} \, , \, \bm{\beta}_{n}}^{(j) \, \mathrm{S}} (\bm{B})$ & $Q_{\bm{\alpha}_{m} \left\{ \bm{\beta}_{n} \right\}_{j}}^{(m+j,m+n)}$ & \\[0.4cm]
 \hline
\end{tabular}
\label{table2}
\end{table}

For clarity, we give in table~\ref{table2} the correspondence between the symmetric quantities analyzed here (for a nonzero magnetic field) and the related quantities considered in \cite{AG07} (where $\bm{B}=\bm{0}$). First of all, we point out two typographical errors that appear in the equations (67) and (68) in \cite{AG07}. Indeed, the last term in the right-hand side of (67) should be $\gamma_{n-m-1} Q_{\alpha_{1} \cdots \alpha_{m} \left\{ \beta \cdots \sigma \right\}_{n-m-1}}^{(n-1,n)}$, while the last term in the right-hand side of (68) should be $\gamma_{n-m} Q_{\alpha_{1} \cdots \alpha_{m} \beta \cdots \sigma}^{(n,n)}$. Taking this remark into account, we see that (i) setting $n=\mathcal{N}$ and $m=\mathcal{N}-2l$, with $1 \leqslant l \leqslant (\mathcal{N}-1)/2$, into the equation (67) of \cite{AG07}, and (ii) setting $n=\mathcal{N}$ and $m=\mathcal{N}-2l+1$, with $1 \leqslant l \leqslant \mathcal{N}/2$, into the equation (68) of \cite{AG07} yields, in view of our notation~\eref{alpha_beta_def} and the correspondence outlined in table~\ref{table2},
\begin{eqnarray}
\fl Q_{\bm{\alpha}_{2\mathcal{N}-2l-2\mathbb{E} \left[ (\mathcal{N}+1) / 2 \right] + 1} \, , \, \bm{\beta}_{2l+2\mathbb{E} \left[ (\mathcal{N}+1) / 2 \right]-\mathcal{N} - 1}}^{\mathrm{S}} (\bm{B}) \nonumber \\[0.2cm]
= \sum_{j=1}^{l} \gamma_{2j-1} Q_{\bm{\alpha}_{2\mathcal{N}-2l-2\mathbb{E} \left[ (\mathcal{N}+1) / 2 \right] + 1} \, , \, \bm{\beta}_{2l+2\mathbb{E} \left[ (\mathcal{N}+1) / 2 \right]-\mathcal{N} - 1}}^{(2j-1) \, \mathrm{S}} (\bm{B})
\label{odd_cumul_sym_expr_from_AG07}
\end{eqnarray}
where the quantities $\gamma_{2j-1}$ are defined (i) by $\gamma_{1}=-1/2$, and (ii) by the recurrence (see equation (59) in \cite{AG07})
\begin{eqnarray}
\gamma_{2j-1} \equiv \frac{1}{4} - \frac{1}{2} \sum_{i=2}^{j-1} \binom{2j-2}{2i-2} \gamma_{2i-1}
\label{gamma_def_AG07}
\end{eqnarray}
for any integer $j \geqslant 2$. Comparing the two expressions~\eref{odd_cumul_sym_expr} and~\eref{odd_cumul_sym_expr_from_AG07} readily shows that
\begin{eqnarray}
\gamma_{2j-1} = e_{0}^{(2j-1)}
\label{gamma_equal_e0}
\end{eqnarray}
for any $j \geqslant 1$. Therefore, substituting~\eref{gamma_equal_e0} into the definition~\eref{gamma_def_AG07}, we obtain the following general identity satisfied by the constant terms $e_0$ of Euler polynomials:
\begin{eqnarray}
e_{0}^{(2j-1)} = \frac{1}{4} - \frac{1}{2} \sum_{i=2}^{j-1} \binom{2j-2}{2i-2} e_{0}^{(2i-1)}
\label{general_identity_e0}
\end{eqnarray}
for any integer $j \geqslant 2$.

We conclude this paper by using results obtained in this section, namely theorems~\ref{theorem2_antisym} and~\ref{theorem2_sym}, to rewrite the quantities $Q_{\bm{\alpha}_{m} \, , \, \bm{\beta}_{n}} (\bm{B})$ in view of the FR~\eref{fluct_rel}. We recall that these quantities are related to their symmetric and antisymmetric parts through~\eref{fct_from_sym_antisym}, i.e.
\begin{eqnarray}
Q_{\bm{\alpha}_{m} \, , \, \bm{\beta}_{n}} \left( \pm \bm{B} \right) = Q_{\bm{\alpha}_{m} \, , \, \bm{\beta}_{n}}^{\mathrm{S}} (\bm{B}) \pm Q_{\bm{\alpha}_{m} \, , \, \bm{\beta}_{n}}^{\mathrm{A}} (\bm{B}) \, .
\label{cumul_from_sym_antisym_def}
\end{eqnarray}
Therefore, for an arbitrary total number $\mathcal{N} \geqslant 1$ of subscripts, we see from~\eref{even_cumul_antisym_expr} that the response of any even cumulant can be written in the form
\begin{eqnarray}
\fl Q_{\bm{\alpha}_{2\mathcal{N}-2l-2\mathbb{E} \left( \mathcal{N} / 2 \right)} \, , \, \bm{\beta}_{2l+2\mathbb{E} \left( \mathcal{N} / 2 \right)-\mathcal{N}}} (\bm{B}) = Q_{\bm{\alpha}_{2\mathcal{N}-2l-2\mathbb{E} \left( \mathcal{N} / 2 \right)} \, , \, \bm{\beta}_{2l+2\mathbb{E} \left( \mathcal{N} / 2 \right)-\mathcal{N}}}^{\mathrm{S}} (\bm{B}) \nonumber \\[0.2cm]
+ \sum_{j=1}^{l} e_{0}^{(2j-1)} Q_{\bm{\alpha}_{2\mathcal{N}-2l-2\mathbb{E} \left( \mathcal{N} / 2 \right)} \, , \, \bm{\beta}_{2l+2\mathbb{E} \left( \mathcal{N} / 2 \right)-\mathcal{N}}}^{(2j-1) \, \mathrm{A}} (\bm{B})
\label{even_cumul_expr}
\end{eqnarray}
where $l$ is any integer such that $1 \leqslant l \leqslant \mathbb{E} \left[ (\mathcal{N}+1) / 2 \right]$. On the other hand, it is clear from~\eref{odd_cumul_sym_expr} that the response of any odd cumulant reads
\begin{eqnarray}
\fl Q_{\bm{\alpha}_{2\mathcal{N}-2l-2\mathbb{E} \left[ (\mathcal{N}+1) / 2 \right] + 1} \, , \, \bm{\beta}_{2l+2\mathbb{E} \left[ (\mathcal{N}+1) / 2 \right]-\mathcal{N} - 1}} (\bm{B}) = Q_{\bm{\alpha}_{2\mathcal{N}-2l-2\mathbb{E} \left[ (\mathcal{N}+1) / 2 \right] + 1} \, , \, \bm{\beta}_{2l+2\mathbb{E} \left[ (\mathcal{N}+1) / 2 \right]-\mathcal{N} - 1}}^{\mathrm{A}} (\bm{B}) \nonumber \\[0.2cm]
+ \sum_{j=1}^{l} e_{0}^{(2j-1)} Q_{\bm{\alpha}_{2\mathcal{N}-2l-2\mathbb{E} \left[ (\mathcal{N}+1) / 2 \right] + 1} \, , \, \bm{\beta}_{2l+2\mathbb{E} \left[ (\mathcal{N}+1) / 2 \right]-\mathcal{N} - 1}}^{(2j-1) \, \mathrm{S}} (\bm{B})
\label{odd_cumul_expr}
\end{eqnarray}
for any integer $l$ such that $1 \leqslant l \leqslant \mathbb{E} \left( \mathcal{N} / 2 \right)$. The two expressions~\eref{even_cumul_expr} and~\eref{odd_cumul_expr} show that the response of \textit{any} cumulant is fully determined once the symmetric part of the response of any even cumulant and the antisymmetric part of the response of any odd cumulant are specified. This reduction is a direct consequence of the FR~\eref{fluct_rel}, and hence of microreversibility.


\section{Conclusion}\label{conclusion_sec}

In this paper, we investigated the consequences of microreversibility on the linear and nonlinear transport properties of general nonequilibrium systems subjected to external magnetic fields. This was done by means of a detailed analysis of the mathematical structure of the multivariate fluctuation relation (FR)~\eref{fluct_rel} satisfied by the generating function $Q \left( \bm{\lambda} , \bm{A} ; \bm{B} \right)$ of the statistical cumulants. The latter fully characterizes the statistics of the currents that flow across the system in a long-time nonequilibrium steady state. We showed that microreversibility imposes constraints on this statistics by unambiguously identifying the independent quantities that are left unspecified by the FR~\eref{fluct_rel}.

The quantities of interest are the statistical cumulants and their responses to the affinities. The $m$th cumulant $Q_{\alpha_1 \cdots \alpha_m} \left( \bm{A} ; \bm{B} \right)$ is obtained upon differentiating the generating function with respect to the counting parameters $\lambda_{\alpha_1}$, \ldots , $\lambda_{\alpha_m}$. Differentiating this cumulant with respect to the affinities $A_{\beta_1}$, \ldots , $A_{\beta_n}$ then yields the $n$th response $Q_{\alpha_1 \cdots \alpha_m \, , \, \beta_1 \cdots \beta_n} (\bm{B})$ of the $m$th cumulant. 

We first derived the set of relations~\eref{gen_rel_cumul_resp} satisfied by the quantities $Q_{\alpha_1 \cdots \alpha_m \, , \, \beta_1 \cdots \beta_n} (\bm{B})$ after expanding both sides of the FR~\eref{fluct_rel} as power series of both the counting parameters $\bm{\lambda}$ and the affinities $\bm{A}$. We then investigated all relations obtained from~\eref{gen_rel_cumul_resp} for a fixed total number $\mathcal{N} = m+n$ of subscripts and all possible values of the indices $m$ and $n$ that are compatible with the constraint that their sum is fixed. After illustrating the relations~\eref{gen_rel_cumul_resp} on the simple cases $\mathcal{N}=2,3,4$, we quantitatively studied the general mathematical structure of~\eref{gen_rel_cumul_resp} for an arbitrary total number $\mathcal{N} \geqslant 1$ of subscripts.

This could be adequately done by decomposing the quantities $Q_{\alpha_1 \cdots \alpha_m \, , \, \beta_1 \cdots \beta_n} (\bm{B})$ in~\eref{gen_rel_cumul_resp} into quantities $Q_{\alpha_1 \cdots \alpha_m \, , \, \beta_1 \cdots \beta_n}^{\mathrm{S},\mathrm{A}} (\bm{B})$ that are symmetric (superscript $\mathrm{S}$) and antisymmetric (superscript $\mathrm{A}$) with respect to the magnetic field. We hence divided~\eref{gen_rel_cumul_resp} into the two distinct sets~\eref{gen_rel_sym_parts} and~\eref{gen_rel_antisym_parts} of relations for the symmetric and antisymmetric parts, respectively. We then separately considered all the relations obtained from~\eref{gen_rel_sym_parts} and~\eref{gen_rel_antisym_parts} for a fixed total number $\mathcal{N} = m+n$ of subscripts.

We first showed that, among all the antisymmetric relations~\eref{gen_rel_antisym_parts}, the only independent ones are obtained from~\eref{gen_rel_antisym_parts} for any even value of the index $m$. This extends the analysis of \cite{BG18} (where no magnetic field is considered) to the antisymmetric relations~\eref{gen_rel_antisym_parts}. An immediate consequence is that the FR~\eref{fluct_rel} leaves the antisymmetric part of any odd cumulant and its responses unspecified, the latter being thus the independent antisymmetric quantities. We then explicitly expressed the remaining dependent quantities, i.e. the antisymmetric part of any even cumulant and its responses, as a linear combination of the independent antisymmetric quantities. This linear combination involves coefficients of Euler polynomials, and generalizes the results of \cite{AG07} (obtained in the absence of a magnetic field) to the antisymmetric quantities $Q_{\alpha_1 \cdots \alpha_m \, , \, \beta_1 \cdots \beta_n}^{\mathrm{A}} (\bm{B})$. Finally, noting that the symmetric relations~\eref{gen_rel_sym_parts} have the exact same form as in the case of a zero magnetic field allowed us to immediately apply the conclusions of \cite{AG07,BG18} to the symmetric quantities $Q_{\alpha_1 \cdots \alpha_m \, , \, \beta_1 \cdots \beta_n}^{\mathrm{S}} (\bm{B})$.

We think that our work contributes to better understand the precise impact of microreversibility on the transport properties of general nonequilibrium systems in the presence of magnetic fields. We emphasize that our results are entirely based on the FR~\eref{fluct_rel}. As such they are hence general consequences of microreversibility, and are in particular independent of the specific underlying microscopic details. It would thus be for instance interesting to illustrate and test the above conclusions on definite models that are known to satisfy a FR of the form~\eref{fluct_rel}. To this end, models that describe mesoscopic transport through quantum dots or Aharonov-Bohm rings (see e.g. \cite{SU08,US09}) are expected to provide a relevant avenue.


\section*{Acknowledgments}

This research is financially supported by the Universit\'e Libre de Bruxelles (ULB) and the Fonds de la Recherche Scientifique~-~FNRS under the Grant PDR~T.0094.16 for the project ``SYMSTATPHYS".


\appendix


\section{Proof of the relation \texorpdfstring{\eref{gen_rel_cumul_resp}}{}}\label{proof_rel_app}

Here we briefly discuss how the general relation~\eref{gen_rel_cumul_resp} can be inferred from the fluctuation relation (FR)~\eref{fluct_rel}. In spite of the presence of an external magnetic field in the present context, the derivation of~\eref{gen_rel_cumul_resp} follows the exact same lines as the corresponding proof presented in the appendix of \cite{BG18} (where no magnetic field is considered). Therefore, we merely restrict our attention on the key steps, and refer the interested reader to \cite{BG18} for additional details.

We first rewrite the FR~\eref{fluct_rel} by making the substitutions $\bm{\lambda}\to-\bm{\lambda}$ and $\bm{B}\to-\bm{B}$ to get
\begin{eqnarray}
Q \left( -\bm{\lambda} , \bm{A} ; -\bm{B} \right) = Q \left( \bm{\lambda} + \bm{A} , \bm{A} ; \bm{B} \right) = \hat T \left( \bm{A} \right) Q \left( \bm{\lambda} , \bm{A} ; \bm{B} \right)
\label{fluct_rel_alt}
\end{eqnarray}
in terms of the translation operator~\eref{transl_op_def}. We then expand both sides of~\eref{fluct_rel_alt} as power series of both the counting parameters $\bm{\lambda}$ and the affinities $\bm{A}$. The left-hand side of~\eref{fluct_rel_alt} is readily written as a power series by means of~\eref{Q_exp_count_par_and_aff}. Thereafter, (i) we expand the translation operator $\hat T \left( \bm{A} \right)$, and (ii) we evaluate the action of the operator $\left( \bm{A} \cdot \partial / \partial \bm{\lambda} \right)^k$, for any integer $k \geqslant 0$, on the power series~\eref{Q_exp_count_par_and_aff} of the function $Q \left( \bm{\lambda} , \bm{A} ; \bm{B} \right)$ to finally get the power series of the right-hand side of~\eref{fluct_rel_alt}. This leads to the following form of the FR~\eref{fluct_rel_alt}:
\begin{eqnarray}
\fl \sum_{m = 0}^{\infty} \frac{(-1)^m}{m!} \sum_{n = 0}^{\infty} \frac{1}{n!} \, Q_{\alpha_1 \cdots \alpha_m \, , \, \beta_1 \cdots \beta_n} (-\bm{B}) \lambda_{\alpha_1} \cdots \lambda_{\alpha_m} A_{\beta_1} \cdots A_{\beta_n} \nonumber\\[0.1cm]
= \sum_{m = 0}^{\infty} \frac{1}{m!} \sum_{k = 0}^{\infty} \frac{1}{k!} \sum_{n=k}^{\infty} \frac{1}{n!} \, Q_{\alpha_1 \cdots \alpha_m \, , \, \beta_1 \cdots \beta_n}^{\{k\}} (\bm{B}) \lambda_{\alpha_1} \cdots \lambda_{\alpha_m} A_{\beta_1} \cdots A_{\beta_n}
\label{fluct_rel_power_series0}
\end{eqnarray}
where the quantities $Q^{\{k\}}$ are defined by $Q^{\{0\}} \equiv Q$ and
\begin{eqnarray}
Q_{\alpha_1 \cdots \alpha_m \, , \, \beta_1 \cdots \beta_n}^{\{k\}} (\bm{B}) \equiv \sum_{j_1 = 1}^{n} \sum_{j_{2}=1 \atop j_{2} \neq j_{1}}^{n} \cdots \sum_{j_{k}=1 \atop j_{k} \neq j_{k-1}}^{n} Q_{\alpha_1 \cdots \alpha_m \beta_{j_1} \cdots \beta_{j_k} \, , \, (\boldsymbol{\cdot})} (\bm{B})
\label{Q_k_def}
\end{eqnarray}
for $k \geqslant 1$, using the notation $(\boldsymbol{\cdot})$ to denote the set of all subscripts $\beta$ that are different to the subscripts $\beta$ present on the left of the comma, i.e. $\beta_{j_1} , \ldots , \beta_{j_k}$ here.

Now, the relation~\eref{fluct_rel_power_series0} must be valid for any counting parameters $\bm{\lambda}$ and affinities~$\bm{A}$. The strategy is thus to identify on both sides of~\eref{fluct_rel_power_series0} the coefficients corresponding to a same power of both the counting parameters and the affinities. The coefficient of the $m$th power of the counting parameters, for any $m \geqslant 0$, is readily obtained, as the sums over the index $m$ are the same on both sides of~\eref{fluct_rel_power_series0}. On the other hand, to identify the $n$th power of the affinities, for any $n \geqslant 0$, first requires to rewrite the two series over the indices $k$ and $n$ in the right-hand side of~\eref{fluct_rel_power_series0} as a single series over an index $n$. We can show that~\eref{fluct_rel_power_series0} is then equivalent to
\begin{eqnarray}
\fl (-1)^m \sum_{n = 0}^{\infty} \frac{1}{n!} \, Q_{\alpha_1 \cdots \alpha_m \, , \, \beta_1 \cdots \beta_n} (-\bm{B}) A_{\beta_1} \cdots A_{\beta_n} \nonumber\\[0.1cm]
= \sum_{n = 0}^{N} \frac{1}{n!} \sum_{j=0}^{n} \frac{1}{j!} \, Q_{\alpha_1 \cdots \alpha_m \, , \, \beta_1 \cdots \beta_n}^{\{j\}} (\bm{B}) A_{\beta_1} \cdots A_{\beta_n} + \mathcal{S}_{N+1}^{\geqslant}
\label{coef_m_power_count_alt_form}
\end{eqnarray}
for any $\bm{A}$, $m \geqslant 0$, $N \geqslant 1$, and where the quantity $\mathcal{S}_{N+1}^{\geqslant}$ only contains powers of the affinities that are at least of degree $N+1$. The fact that the integer $N$ is arbitrary allows us to identify on both sides of~\eref{coef_m_power_count_alt_form} the coefficients of any power $n \geqslant 0$ of the affinities. Finally, noting that the quantity $Q^{\{j\}}$ as defined by~\eref{Q_k_def} can, in view of the invariance~\eref{Q_inv_perm_indices}, be written as
\begin{eqnarray}
Q_{\alpha_1 \cdots \alpha_m \, , \, \beta_1 \cdots \beta_n}^{\{j\}} (\bm{B}) = j! \, Q_{\alpha_1 \cdots \alpha_m \, , \, \beta_1 \cdots \beta_n}^{(j)} (\bm{B})
\label{Q_j_alt_expr}
\end{eqnarray}
in terms of the quantities $Q^{(j)}$ defined by~\eref{Q_j_expr}, we get the desired result~\eref{gen_rel_cumul_resp} upon multiplying both sides of~\eref{coef_m_power_count_alt_form} by $(-1)^m$ and reversing the magnetic field ($\bm{B}\to-\bm{B}$). Q.E.D.


\section*{References}

\bibliographystyle{unsrt}
\bibliography{DATABASE_Nonlinear_response_with_B}


\end{document}